# Topological Textures in Zr-Substituted Barium Titanate

Florian Mayer[1*]

[1]Materials Center Leoben Forschung GmbH, Vordernberger Strasse 12, 8700 Leoben, Austria

*florian.mayer@mcl.at

## Abstract

Topological polarization textures in ferroelectrics offer pathways to dense memory, neuromorphic computing, and controlled probes of topology in solids. In rhombohedral barium titanate, theory has identified stable antiskyrmions of topological charge −2 that fractionalize into six −1/3 hotspots, termed topological quarks. Here we extend this landscape to Zr-substituted barium titanate (BZT) using a first-principles parameterized effective Hamiltonian framework. In an ordered 12.5% composition, the chemically doubled periodicity enforces an alternation along [111]: one half hosts the −2 antiskyrmion (six −1/3 quarks), the other a +4 skyrmion (six +2/3 quarks). The two share the same six-vortex scaffold (threefold motif) but differ in the core-level polarization texture, resulting in an integer +1 per vortex offset in the plane-integrated (slice) topological charge. In random BZT, nanodomains remain inducible and cryogenically stable, yet quenched disorder pins and distorts the vortices, producing a heterogeneous, skyrmion-glass-like state with fluctuations of the topological charge along the nanodomain axis. Thermal stability maps show that pure barium titanate retains −2 textures up to ~100 K, whereas in BZT the collapse temperature is nonmonotonic, with a minimum near 6–8% Zr, reflecting competition between ferroelectric softening and disorder pinning. Importantly, the 12.5% ordered arrangement remains rhombohedral above 300 K, enabling field-stabilized nanodomains at 293 K. Under a local [111] bias, the ordered system carries +4 slice charge, while the random composition fragments under the same conditions. These results establish BZT as a platform for chemically programmed, fractionalized ferroelectric topology from cryogenic to room temperature and suggest routes to multistate, reconfigurable devices.

# I. INTRODUCTION

The study of topological quasiparticles [1–4] has become a rapidly growing branch of condensed-matter physics because these excitations combine nanoscale stability with rich emergent electrodynamics [5]. Early research concentrated almost exclusively on magnetic topological textures [6–8], where the antisymmetric Dzyaloshinskii–Moriya interaction (DMI) is the key ingredient. More recently, attention has shifted to polar textures that arise in ferroelectric crystals [9–11]. In such systems the driving forces are the anisotropy of the local electric dipoles and their coupling to lattice distortions, rather than DMI. Polar skyrmions were first identified in ferroelectric/paraelectric multilayers [10,11], following the observation of polar vortices in $PbTiO_3/SrTiO_3$ superlattices [12,13]. Since then, an increasing number of purely ferroelectric materials have now been predicted, or even observed, to harbor non-trivial topology [14–20]. Prominent examples include "skyrmion bubbles'' in $PbTiO_3$ [18] obtained from molecular-dynamics (MD) simulations. Complementary progress includes the discovery of emergent topological polarization textures (vortices/merons) in lead-based relaxor ferroelectrics such as PMN-PT [21] and the experimental imaging of meron–antimeron networks in moiré ferroelectrics based on twisted vdW bilayers [22]. Especially intriguing for the present work is the theoretical prediction that bulk barium titanate ($BaTiO_3$, BT) in its rhombohedral phase can sustain a variety of polar textures. Gonçalves et al. [20] showed that antiskyrmions with a topological charge of –2 can be created by locally inverting the polarization. In subsequent studies it was demonstrated that the same objects can be written and steered with external electric fields [14], and that one-dimensional (1D) line defects [17] (skyrme lines) are also possible. Although most -2 antiskyrmion calculations were performed at cryogenic temperatures, previous work [15] revealed theoretical thermal stability up to ~85 K. More recently, Gómez-Ortiz et al. [16] proposed electrically switchable skyrmion–antiskyrmion tubes with $Q=\pm 1$ in rhombohedral BT. Beyond electric-field control, structured light has recently been proposed to manipulate ferroelectric topologies, including dynamical polar (anti)skyrmions and chirality selection [23].

Beyond their mere existence, the internal structure of the –2 antiskyrmions is remarkable. For diameters below ~4 nm the topological charge density is fragmented into six hotspots carrying –1/3 each, entities denoted as topological quarks [15,20]. They exhibit well-defined positions, charges and quasiparticle-like behavior. When the diameter exceeds ~4.5 nm every –1/3 quark splits into two –1/6 pre-quarks while the global symmetry and net charge are preserved [15]. The non-trivial topology is thus concentrated at the antiskyrmion vertices, whereas the regions in between relax into a trivial 180° domain wall [15]. These observations already portray BT as an ideal playground for exploring how topology evolves with size and dimensionality. Topological objects with $|Q|>1$ are not merely academic curiosities [24–26]. A larger net charge magnifies the emergent electromagnetic fields and promises stronger electro-optic responses as well as genuine multi-level states for non-volatile memories. Hosting positive and negative charges in the same crystal would further allow domain-engineered conduits for topological currents and offer a solid-state analogue of confinement phenomena usually discussed in high-energy physics. These considerations motivate the search for systems in which possible skyrmion and antiskyrmion textures can coexist and interact.

With this objective in mind, and motivated by the already rich set of predicted topological textures in pure BT, we extend our study to investigate the effects of substituting Ti with Zr, i.e., barium zirconate titanate ($\mathrm{Ba(Zr_xTi_{1-x})O_3}$, BZT). The motivation arises from the BZT phase diagram (Refs. [27–29]), which shows that with an appropriate choice of Zr concentration, the rhombohedral phase can be stabilized above room temperature. This rhombohedral structure is precisely the prerequisite for the topological textures previously induced in pure BT [14,15,20]. The relevant concentration range for this large stabilizing effect of the rhombohedral phase lies around 10–15% Zr [27]. In this study, we carry out finite-temperature molecular dynamics (MD) simulations using a first-principles parameterized effective Hamiltonian [27] for BZT, computational details are provided in the next section. First, we study a high-symmetry ordered 12.5% Zr arrangement in which the B-site lattice is doubled along each Cartesian direction (period two), yielding one Zr per doubled cell and a 1/8 overall substitution. In this system, we again induce nanodomains [15,20] and find that stable topological textures persist. Remarkably, due to the doubling of the unit cell introduced by chemical substitution, the well-known charge −2 antiskyrmion [14,15,20] undergoes a splitting. More specifically, the doubled unit cell separates into two topologically distinct regions: one half hosts the previously reported −2 antiskyrmion composed of six −1/3 topological quarks, while the other half hosts a +4 skyrmion with six +2/3 quarks. This behavior mainly originates from the underlying chemical sublattice and the accompanying change in local anisotropy. In addition to the real-space analysis, we introduce an optional symmetry-based bookkeeping perspective for the chemically doubled stacking and show that the +1-per-vortex offset is mediated by localized interfacial core-conversion events. Next, we examine whether stable topological textures can also be sustained in systems with randomly distributed Zr ions. We find that nontrivial topological textures remain robust under random substitution of Zr. However, with increasing Zr concentration, these textures become distorted and deviate from the highly symmetric configuration. All of these simulations are carried out at low temperature (1 K). We then address the intrinsic thermal stability of these textures and find that, depending on concentration, they can remain stable up to nearly 100 K. Finally, we take one step further and demonstrate that in the 12.5% superlattice, topological textures can also be induced near room temperature by applying external electric fields, although their stability requires the continued presence of the field.

Overall, this work presents a comprehensive study of topological textures in Zr-substituted BT. We show that chemical substitution not only preserves but also enriches the zoo of emergent textures by enabling unit-cell doubling, which gives rise to an interplay of fractionalized topological charges (−1/3 and +2/3 quarks). We further demonstrate that these textures remain robust under random substitution, can be stabilized up to nearly 100 K, and, with the aid of external electric fields, can even be induced at room temperature.

# II. COMPUTATIONAL DETAILS

In this study, the computational experiments were performed using MD simulations based on effective Hamiltonians (EH) [28,30–36]. Effective Hamiltonians provide the advantage of being highly efficient [37] while still delivering a reliable, quantitative description of the underlying atomistic effects [36,38]. The parametrizations employed here were developed in the works of Mayer et al. [27,38], incorporating anharmonic couplings for pure BT as well as an improved treatment of substitution effects. More details on the Hamiltonian for pure BT can be found in Ref. [38], while the Hamiltonian applied for Zr-substituted BT is comprehensively described in Ref. [27]. For completeness, a summary of the model and notation is provided in the Supplemental Material [39]. Within this EH framework, the microscopic order parameter is a B-site–centered local mode vector $\boldsymbol{u}(\boldsymbol{R})$ defined for each 5-atom perovskite unit cell. It is directly associated with the corresponding local dipole moment $\boldsymbol{d}(\boldsymbol{R}) = Z^*(\sigma_{\boldsymbol{R}})\boldsymbol{u}(\boldsymbol{R})$, where $Z^*(\sigma_{\boldsymbol{R}})$ is the (species-dependent) mode Born effective charge [27,39] of the B-centered unit cell at site R. The local polarization is then $\boldsymbol{p}(\boldsymbol{R}) = \boldsymbol{d}(\boldsymbol{R})/\Omega_0$, with $\Omega_0$ the unit-cell volume. For topology we use the normalized order-parameter field $\boldsymbol{n}(\boldsymbol{R}) = \boldsymbol{u}(\boldsymbol{R})/|\boldsymbol{u}(\boldsymbol{R})|$ (equivalently $\boldsymbol{n}(\boldsymbol{R}) = \boldsymbol{p}(\boldsymbol{R})/|\boldsymbol{p}(\boldsymbol{R})|$). Another clear advantage of these parametrizations is their accurate description of phase transition temperatures for both pure and substituted BT [27,38], which ensures meaningful and reliable simulation results. All simulations were carried out using a modified version of the *feram* software [34,35]. The simulations were conducted in the canonical ensemble, with temperature scaling achieved using a velocity-scaling algorithm, validated against Nosé–Poincaré [40] results. A timestep of 2 fs was employed for all simulations. Each trajectory ran for 1 ns (800 ps equilibration + 200 ps production). The simulation cell was fixed at 64×64×64 unit cells with periodic boundary conditions in all three directions employed. For disordered BZT, we performed three independent simulations with randomized B-site occupations to assess statistical variability and mitigate seed-specific effects.

For characterizing topological textures within our study, we utilize the Pontryagin density (see Equation (1)) which quantifies the topological charge for a continuum unit-vector order-parameter field $\boldsymbol{n}(\boldsymbol{r})$ defined over a two-dimensional spatial domain (compactified to $S^2$) and mapping into the two-sphere target space $S^2$, corresponding to the homotopy group $\pi_2(S^2) = \mathbb{Z}$.

For the actual computation of the topological charges $Q$ we use the lattice formulation proposed by Berg and Lüscher [41], as illustrated in the work of Heo et al. [42]. In the latter work, the lattice discretization is given in Equations (2) and (3). The summation is carried out over the areas $A_l$ of the spherical triangles spanned by the vectors $\boldsymbol{n}_i$, $\boldsymbol{n}_j$, and $\boldsymbol{n}_k$, where $\boldsymbol{n}_i \equiv \boldsymbol{n}(\boldsymbol{R}_i)$ denotes the normalized order-parameter field on lattice site $i$. Here, the spin vectors of Ref. [42] are replaced by the corresponding normalized vectors constructed from the effective-Hamiltonian local modes (or equivalently the local polarization field). The sign is determined by the relation $sign(A_l) = sign[\boldsymbol{n}_i \cdot (\boldsymbol{n}_j \times \boldsymbol{n}_k)]$.

$$Q = \frac{1}{4\pi} \int_{\Omega} \boldsymbol{n} \cdot (\partial_x \boldsymbol{n} \times \partial_y \boldsymbol{n}) d^2 r \tag{1}$$

$$Q = \frac{1}{4\pi} \sum_{l} A_l \tag{2}$$

$$\cos\left(\frac{A_l}{2}\right) = \frac{1 + \boldsymbol{n}_i \cdot \boldsymbol{n}_j + \boldsymbol{n}_j \cdot \boldsymbol{n}_k + \boldsymbol{n}_k \cdot \boldsymbol{n}_i}{\sqrt{2(1 + \boldsymbol{n}_i \cdot \boldsymbol{n}_j)(1 + \boldsymbol{n}_j \cdot \boldsymbol{n}_k)(1 + \boldsymbol{n}_k \cdot \boldsymbol{n}_i)}} \tag{3}$$

# III. RESULTS

## A. Alternating -2/+4 Textures in a Chemically Doubled Unit Cell

Since the existence of −2 antiskyrmions in BT relies on the rhombohedral phase [20] (R3m), we begin by constructing a system that preserves the crystallographic prerequisites for such textures and thus may also host nontrivial topology. For Zr-substituted BT we must remain in the concentration range where the rhombohedral phase is stable. The BZT phase diagram (Ref. [27–29]) shows that the transition temperature from rhombohedral to orthorhombic symmetry initially increases with Zr concentration, reaching a maximum at about 10–15% where it exceeds 300 K, before decreasing again at higher substitution levels. We therefore choose 12.5% and construct a 2×2×2 cell, where one Ti ion is replaced by Zr, yielding the desired composition. To assess whether the superlattice satisfies the basic prerequisites for hosting topological textures, we performed initial first-principles density-functional theory (DFT) calculations [43–48]. These show that the superlattice adopts a rhombohedral ground state with spontaneous net polarization along [111], providing the crystallographic setting required for the textures reported in pure BT. We also computed the spontaneous polarization and compared it with results from our EH simulations (see Figure S1 [39]). The two approaches exhibit reliable quantitative agreement (within the numerical uncertainty of the EH parametrization). Importantly, despite the presence of nominally nonpolar Zr-centered cells, the superlattice retains a substantial net polarization. Together, these findings indicate that the structural and polar conditions necessary to stabilize topological textures, analogous to those in pure BT, are indeed met in the 12.5% Zr superlattice. More details on these calculations and of this superlattice is provided in the Supplemental Material [39]. With an effective Hamiltonian for BZT available [27], we can now perform large-scale MD simulations [27,37]. We follow the procedure of Ref. [15] and consider a 64×64×64 supercell with Zr distributed as described above. Initially, we restrict ourselves to a low temperature of 1 K. To induce nanodomains, the supercell is first initialized as a single domain with polarization along [111]. We then invert the polarization within approximately cylindrical regions aligned with [111], effectively creating nanodomains with circular cross-section. This serves as the starting configuration: from here, the MD simulation is allowed to evolve freely so that the system can relax. In this superlattice we introduce nanodomains with diameters ranging from ~2.2 to ~10 nm. The system is equilibrated for 800 ps at 1 K, and results are averaged over a subsequent 200 ps. We find that nanodomains with diameters between ~3 and ~10 nm remain stable, while the smallest domain (~2.2 nm) decays into a single-domain state within a few tenths of picoseconds. Thus, as in pure BT, stable nanodomains are formed in the 12.5% BZT superlattice. The central question is now whether these nanodomains also host nontrivial topological textures. To answer this, we analyze cross-sections in planes normal to the nanodomain axis, i.e., the (111) planes. Here, an additional symmetry consideration becomes important. In the undoubled lattice, there are three symmetry-distinct (111) planes per period along [111], analogous to ABC stacking in fcc. This trio constitutes the fundamental geometric unit of the texture. In the present case, chemical substitution doubles the unit cell along [111]. While the geometric stacking remains unchanged, the chemical motif alternates, resulting in two inequivalent halves: one with Ti–Zr–Ti stacking and the other with Zr–Ti–Zr. In other words, the chemical motif is reversed between the two halves of the doubled unit cell. For the analysis of topological textures, we therefore work with the three symmetry-distinct geometric (111) planes that constitute the topological unit, whereas the alternating chemical motif acts as a background modulation and does not redefine these planes. Consequently, unlike in previous studies [15,20], two distinct (111) projections must be considered due to the doubled unit cell along [111].

We now proceed with our initial analysis. Following Section II, we compute the plane-resolved topological charge for projections onto the (111) planes in geometric triplets. Specifically, at each position z′ we define the plane-integrated topological charge (slice charge) as $Q(z') = \int q(x', y'; z') dx' dy'$, where $q(x', y'; z')$ is the local topological charge density on the (111) plane at fixed z′ (see Equation (1) for the Pontryagin density and Equation (3) for its lattice evaluation). We then translate the projection window along the nanodomain axis and repeat. The axial direction z′ is defined by rotating the simulation cell so that z′ is parallel to [111]. The resulting $Q(z')$ is shown in Figure 1a. Each data point corresponds to a single (111) plane, so that three consecutive points form one geometric triplet along z′.

A striking alternation emerges. One half of the chemically doubled unit cell (hatched region) hosts the familiar −2 antiskyrmion known from pure BT [20], whereas the other half hosts a +4 skyrmion. This −2/+4 pattern repeats periodically along the nanodomain (z′) direction. To compare the real-space textures, Figures 1c and 1d display the (111)-plane projections of the two halves using the rotated in-plane axes x′||$[1\bar{1}0]$ and y′||$[11\bar{2}]$. We also plot the local topological charge density $q(x', y'; z')$. Both patterns in Figures 1c and 1d exhibit a hexagonal-like in-plane dipole arrangement on (111), consistent with the parent 3m point-group symmetry [15,20]. In Figure 1d, the −2 antiskyrmion (diameter ~3 nm) comprises six vortices with alternating clockwise and counter-clockwise circulation, together with three vertices where the in-plane polarization diverges outward and three where it converges inward, fully consistent with pure BT [14,15,20]. Each vortex coincides with a hotspot of $q(x', y'; z')$ carrying approximately −1/3, yielding six −1/3 topological quarks. By contrast, Figure 1c shows that the +4 texture retains the same six-vortex scaffold (threefold motif), but the polarization texture around the individual cores differs measurably from the −2 half, interestingly, the hotspots now carry +2/3. The −1/3 and +2/3 hotspots occur in the same six angular sectors (see Figures 1c-1d and Figure S2 in the Supplemental Material [39]), indicating that the global six-vortex scaffold and 3m symmetry are preserved, while the core-level geometry differs and the fractional charges shift.

We can interpret these results as follows: The alternation of −2 and +4 textures originates from the two-sublattice chemical motif introduced by Zr substitution, which doubles the unit cell along [111]. Although the geometric stacking of planes is unchanged, the chemical motif is reversed between the two halves (Ti–Zr–Ti ↔ Zr–Ti–Zr). This reversion modifies the local anisotropy landscape and, effectively, the coefficients of the gradient-like terms (in the EH sense of composition-dependent short-range couplings [27]) on alternating (111) planes. As a result, the plane-resolved skyrmion density acquires an integer offset when passing from one half to the other: each vortex's fractional charge changes from

−1/3 to +2/3, i.e., by +1. Summing over the six vortices therefore shifts the total from −2 to +4 while preserving the texture's morphology: $-2 + 6 \times (+1) = +4$.

The −2/+4 alternation emerges directly from how the slice-resolved skyrmion density redistributes across the chemically doubled cell. In real space, the inter-half difference map $dq(x', y')$ shows localized peaks at the vortex cores while the six-vortex scaffold remains trackable across the interface. Integrating $dq$ over each peak yields an approximately integer +1 per vortex line (total ΔQ≈+6 per half-cell), with vortex positions and overall 3m point group symmetry essentially unchanged (see Supplemental Material [39]). The chemical interface therefore acts as an effective conduit of topological charge that converts the local fractional assignments on each slice (e.g., −1/3 to +2/3 on matching vortex cores) without requiring lateral motion of the cores.

Two immediate consequences follow: (a) the −2/+4 sequence constitutes a topological charge-density wave along z′ whose wavelength is fixed by the chemically doubled periodicity, (b) because the six vortex scaffold is nearly invariant through the interface, the switching of fractional charge is governed primarily by the layer-dependent anisotropy set by the chemical sublattice, offering a route to engineer fractionalization via compositional anisotropy modulation rather than by altering nanodomain geometry.

The splitting of the topological texture described above is observed for nanodomain diameters up to approximately 4.5 nm. For larger diameters, the system exhibits the previously reported splitting [15] of the topological quarks (e.g., from −1/3 into −1/6), and the broadened cores sample multiple low-$|\boldsymbol{p}|$ sites, which blurs the clean −2/+4 alternation and produces more complex patterns that we do not pursue further here. In the next section, we provide an optional symmetry-based bookkeeping perspective for the observed integer offset between the two chemical halves and discuss its microscopic realization via localized core-conversion events.

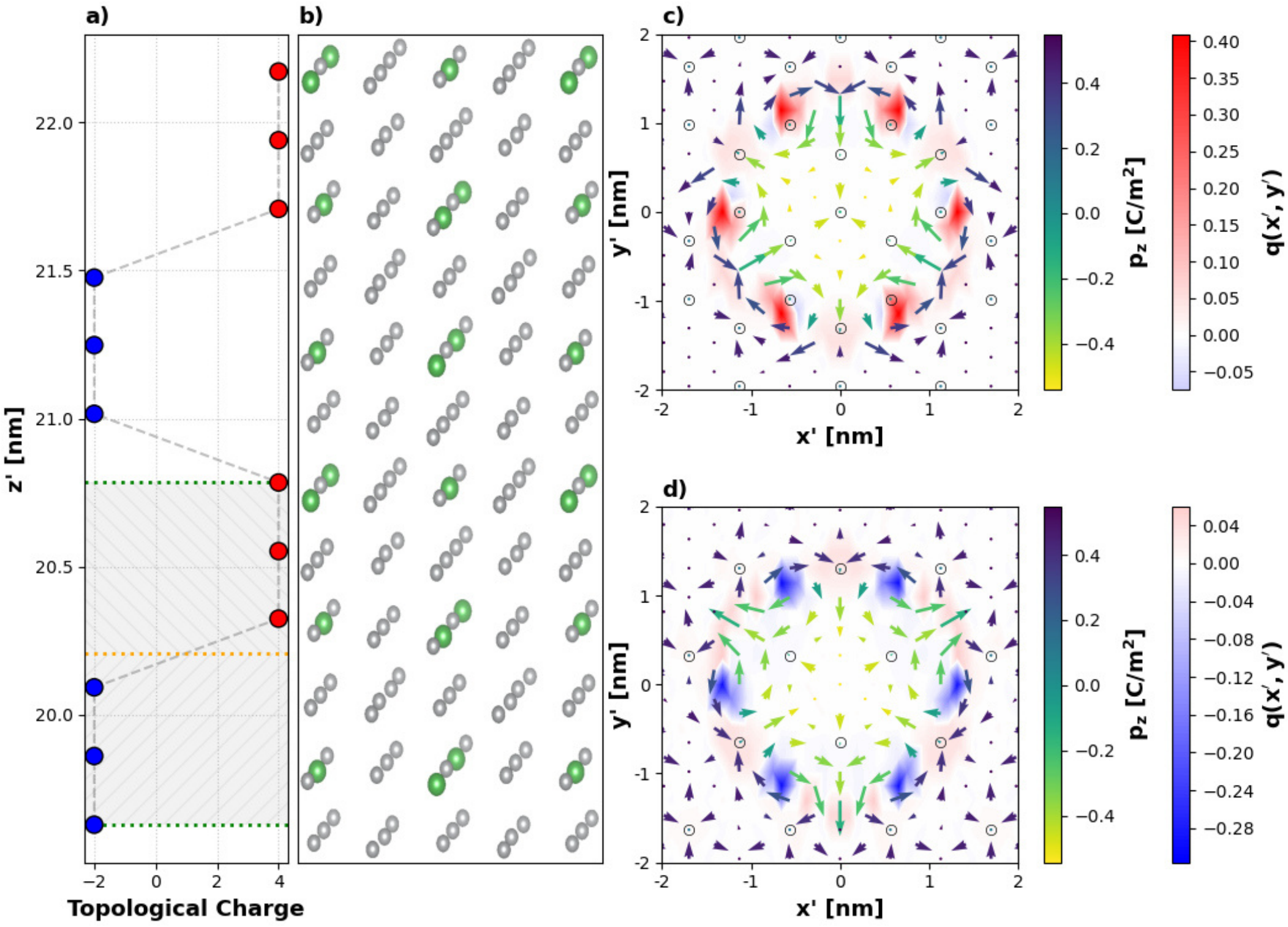

*Figure 1. Topological behavior in the 12.5% superlattice along the nanodomain axis z′. (a) Integrated topological charge along z′∥[111]. The shaded region marks the chemically doubled unit cell, and the orange line indicates the chemical interface associated with the change in topological character. (b) Schematic associated B-site ordering: gray (smaller) spheres indicate Ti ions, and green (larger) spheres represent Zr ions. (c) Local polarization vectors of the Q=+4 skyrmion, colored by the out-of-plane component. The overlaid local topological charge density reveals its fragmentation into six +2/3 topological quarks. (d) The Q=−2 antiskyrmion, similarly decomposed into six −1/3 topological quarks. Panels (c) and (d) are shifted along z′ relative to each other, consistent with the offset seen in panel (a). In the latter panels, small circles indicate Zr substituted unit cells.*

## B. Symmetry Bookkeeping and Microscopic Charge-Transfer Mechanism

As an optional view alongside the real-space narrative, we can model the chemically doubled unit cell along $z' \parallel [111]$ as a static binary background. The two halves (Ti–Zr–Ti and Zr–Ti–Zr) can be considered as an Ising-like "color'' field with $\mathbb{Z}_2$ link variables that are +1 within a half-cell and flip sign across the chemical interface. In this notation, a half-cell translation along z′ corresponds to a fixed sign change, whereas a full-cell shift is ordinary. We emphasize that this construction is optional and does not enter the quantitative analysis of Section III.A, it is offered solely as a compact language for discussing symmetry and phase conventions imposed by the chemical doubling.

In this language, the color-sensitive quantity is a phase-like variable that records the in-plane rotation of the polarization relative to a chemically anchored reference frame. Crossing the chemical interface applies a fixed 180° flip to that reference frame, so this variable changes sign between the two halves, even though the polarization pattern can evolve continuously with the local anisotropy. This background-imposed sign convention partitions adjacent planes into inequivalent bookkeeping sectors and provides a compact description of the observed alternation.

Analyzed plane by plane along z′, one half of the doubled cell carries a plane-integrated topological charge of −2, while the other carries +4. In our topological quark language, the −2 texture decomposes into six −1/3 quarks, whereas the +4 texture decomposes into six +2/3 quarks at essentially the same in-plane positions. The color background provides a compact description: a half-translation corresponds to a fixed $\mathbb{Z}_2$ phase convention (a π reference-frame flip) between the two chemical halves, which provides compact bookkeeping for the observed +1-per-vortex offset. Summing over six aligned vortices shifts −2 to +4 while preserving the six-vortex scaffold (coincident vortex-axis positions), even though the core-resolved polarization textures differ between the two halves.

Regarding how the integer offset is realized microscopically, a change of the plane-resolved skyrmion number requires a breakdown of the normalized order-parameter field $\boldsymbol{n}(\boldsymbol{R})$ (i.e., $\boldsymbol{n}(\boldsymbol{R}) \equiv \boldsymbol{p}(\boldsymbol{R})/|\boldsymbol{p}(\boldsymbol{R})|$, equivalently $\boldsymbol{u}(\boldsymbol{R})/|\boldsymbol{u}(\boldsymbol{R})|$), a singular event in the local-mode texture. In our data the six vortex cores remain approximately axially aligned across the chemical interface, while $|\boldsymbol{p}|$ becomes strongly suppressed at the vortex axis within the interfacial region (Figure S2 and S3 in the Supplemental Material [39]), consistent with Bloch-point–like polarization singularities (local $|\boldsymbol{p}| \to 0$) [49–51]. Microscopically, such nodes are naturally facilitated by Zr-centered cells: compared with Ti-centered environments, Zr substitution yields locally "softer" polar response and can support configurations with very small local polarization amplitude. This provides chemically available sites at which the vortex axis can transiently extinguish $|\boldsymbol{p}|$, enabling the required integer jump in the plane-resolved topological charge without lateral motion of the vortex cores. In this sense, the interface acts as a chemically assisted conversion region, where Bloch-point–like nodes serve as local sources/sinks of topological density and transfer one integer unit per vortex line between adjacent (111) sections. The $\mathbb{Z}_2$ construction above serves as optional bookkeeping for this integer offset, the microscopic conversion is mediated by the localized singular cores described here.

Two internal checks are consistent with this picture: (a) the plane-resolved $Q(z')$ from Figure 1a exhibits a clean −2/+4 alternation within a single nanodomain, (b) subtracting the local topological charge densities of the two halves yields six coincident peaks, each integrating to approximately +1, i.e., one unit gained per quark, under the $\mathbb{Z}_2$ bookkeeping convention. The latter check is provided in Figure S4 in the Supplemental Material [39]. This proposed organizing principle presumes that corresponding vortex cores remain approximately axially aligned along z′ (as observed for diameters below ~ 4.5 nm). Strong disorder or lateral wandering can already blur the alternation. In addition, as the skyrmion diameter increases and the vortex cores broaden, the number of Zr-centered cells sampled within a given core increases, providing multiple low $|\boldsymbol{p}|$ candidate sites for singular conversion events, this can delocalize the charge-transfer process along z′ and degrade the clean −2/+4 alternation within a single nanodomain. In that regime a purely real-space description is more appropriate. Within its range of validity, the color background provides a compact, alternative field-theoretic lens on the observed −2→+4 sequence.

## C. Topological Textures in Randomly Substituted BZT

Having established that nontrivial topological textures can be induced in the highly symmetric 12.5% superlattice, we now turn to a more experimentally accessible scenario: random Zr substitution. Our protocol (nanodomain induction, supercell size, MD settings, and plane-resolved analysis) is identical to the superlattice case. The only change is the composition: we distribute Zr ions at random in steps of 2% from 0% (pure BT) up to 12.5% within the simulation box. We focus on an induced nanodomain of diameter ~4 nm and fix the temperature at 1 K. Overall, for all simulated compositions, nanodomains can be created and remain stable over the full simulation window. Figure 2 summarizes the associated topological textures using a single projection along z′. In the absence of imposed chemical doubling, this serves as a first indicator rather than a full 3D characterization. Nevertheless, because Zr ions are distributed randomly, the textures do show variations along the nanodomain axis. In Figure 2a (0% Zr), pure BT reproduces the canonical −2 antiskyrmion with its six-vortex scaffold and the familiar threefold rotational symmetry. The plane-resolved topological charge remains essentially constant at Q=−2 along the nanodomain. In Figure 2b (2% Zr), the global threefold symmetry is still preserved, but local anisotropy fluctuations introduced by isolated Zr ions distort individual vortices. In the local charge density $q(x', y';\, z')$, this manifests as a positive hotspot at one vortex core. As a result, the plane-integrated charge for that section is reduced to topological charge of −1 instead of −2. This demonstrates that local defects can modify the topological charge while leaving the global six-core scaffold and threefold motif recognizable. At 6% substitution (Figure 2c), disorder effects become more pronounced. Vortex cores exhibit pinning and shear, so the threefold symmetry remains visible at the mesoscopic scale but is clearly broken locally. The charge density $q(x', y';\, z')$ displays both positive and negative hotspots of varying strength. In several regions, we observe partial conversion events in which only subsets of vortices undergo charge transfer, leading to irregular fluctuations of $Q(z')$ (see Figure S5 in the Supplemental Material [39]) along the nanodomain axis. At the highest investigated concentration, 12.5% (Figure 2d), the nanodomain remains remarkably robust, and a nontrivial texture is still sustained. The pattern resembles a patchwork version of the pure case, with a reduced correlation length: multiple hotspots appear at the characteristic scaffold positions, yet the six-vortex framework is still recognizable. The plane-resolved charge $Q(z')$ (see Figure S5 [39]) varies significantly along the axis and frequently deviates from the canonical Q=−2 value of pure BT, indicating that the texture remains topologically nontrivial and strongly modulated by compositional disorder.

We can interpret these results as follows: Random Zr substitution introduces quenched anisotropy and random-field disorder that significantly influence the stability and morphology of the textures. First, local variations in the chemical environment pin and bend vortex lines, leading to distortions of the otherwise symmetric vortex scaffold. Second, these variations generate narrow interfacial regions where the local topological charge density $q(x', y';\, z')$ changes abruptly, thereby enabling charge transfer between neighboring planes (see Supplemental Material [39]). Third, such processes give rise to a spatially heterogeneous redistribution of fractional charges: vortices carrying −1/3 in pure BT can locally convert into +2/3, combine to larger fractions, or split into smaller fractions, depending on the local arrangement of Zr ions. The emergent state can thus be viewed as a skyrmion-glass-like texture pinned by quenched disorder, by analogy with magnetic skyrmion glasses [52,53]. The six-vortex scaffold characteristic of the −2 texture remains broadly recognizable, but its internal fractionalization is strongly modulated by compositional randomness. Because these charge-transfer events occur irregularly along the nanodomain axis, the system cannot be unambiguously classified into a single texture type (e.g., −2 or +4) as in the ordered superlattice. Instead, the textures acquire a patchwork character, with local motifs resembling fragments of the ordered cases embedded in a disordered background.

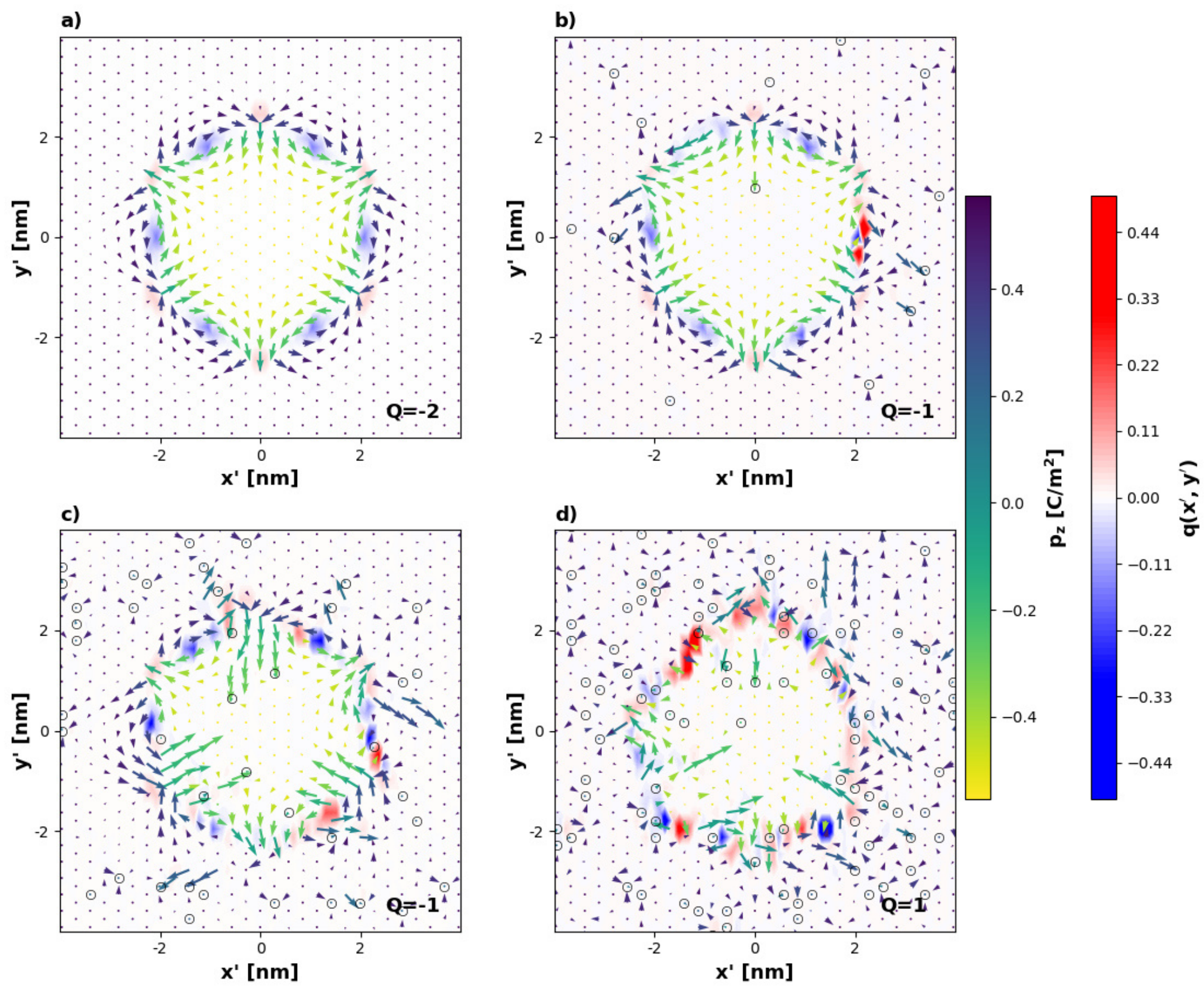

*Figure 2. Topological textures in random BZT. Each panel shows the in-plane polarization vectors in the (111) projection, color-coded by the out-of-plane component, with the local topological charge density overlaid. (a) Pure BT, displaying the canonical Q=−2 antiskyrmion. (b) 2% randomly distributed Zr, showing the same underlying texture with minor defects. (c) 6% randomly distributed Zr, where defects become more pronounced but the overall symmetry is still largely preserved. (d) 12.5% randomly distributed Zr, exhibiting strong deviations and fragmentation of the topological charge into multiple pieces, while still retaining a coarse imprint of the original symmetry. In all panels, small circles indicate Zr substituted unit cells.*

## D. Intrinsic Thermal Stability vs Zr Concentration

So far, our simulations have been performed at very low temperature (1 K), where nanodomains and their associated topological textures are stable. For experimental relevance and prospective applications, we now assess thermal stability. Following Ref. [15], we probe the intrinsic stability by ramping the temperature in MD while monitoring the plane-resolved topological charge Q. Specifically, we initialize three different nanodomains (∼3.5, ∼5, and ∼6.5 nm) at 1 K, then increase the temperature in 5 K steps until collapse. The collapse temperature $T_{collapse}$ is defined as the temperature at which the nanodomain irreversibly collapses into a single-domain state under our simulation protocol. We perform such heating runs for random Zr concentrations from 0% to 12.5% (in 2% steps) and for the 12.5% superlattice. For all presented results, only one (111) projection is used to compute the topological charge in Figure 3a-3c, however, as noted earlier (Figure S5 [39]), randomness of the Zr distribution also induces fluctuations of Q along the core.

We first examine how the plane-resolved topological charge evolves with temperature. As noted above, we consider three nanodomain diameters across random Zr concentrations and in the 12.5% superlattice. Figures 3a–3c summarize these results. For pure BT, all three diameters retain the canonical −2 texture over almost the entire temperature range up to $T_{collapse}$, consistent with Ref. [15]. In the superlattice, the behavior is likewise nearly constant. That means for the 3.5 nm domain the alternating −2/+4 pattern persists also to finite temperature (even up to $T_{collapse}$), whereas for 5 nm and 6.5 nm the alternation has vanished and the topological texture along z′ constantly shows the −2 behavior. (Figure 3a displays only the −2 branch for the superlattice because a single (111) projection along z′ is analyzed.)

For randomly Zr-substituted systems, the behavior changes qualitatively. Already at 2% Zr, fluctuations in the topological charge appear upon heating. Nevertheless, nanodomains remain topologically nontrivial throughout their thermodynamic stability window. We note that Q=0 in a given (111) slice does not necessarily imply a trivial texture, since positive and negative fractional contributions can cancel in the plane integral while a nontrivial vortex scaffold remains present. With increasing Zr content the fluctuations grow, and occasional excursions to more exotic integer values are observed, although such events are rare, instead the predominant values lie between −2 and +2. These trends again indicate that Zr introduces local anisotropy variations that distort and pin the vortex skeleton, producing a heterogeneous, disorder-modulated sequence of plane textures. In short, random substitution induces charge fluctuations upon heating and along z′, yet the nontrivial character persists until the nanodomains eventually collapse.

Figure 3d explicitly illustrates the collapse temperatures as a function of diameter and composition. In detail, for each concentration, we performed three independent heating runs with randomly distributed Zr, that means we report the mean $T_{collapse}$ across these runs and estimate the uncertainty from their spread. For pure BT, increasing diameter modestly enhances stability. That means we obtain a $T_{collapse}$ of ~65 K for 3.5 nm and ~110 K for 6.5 nm (the 5 nm case lies in between), in agreement with Ref. [15]. With random Zr the $T_{collapse}$ decreases for all diameters, most strongly for 3.5 nm and less strongly as the diameter grows. A minimum occurs near 6–8% Zr, after which $T_{collapse}$ recovers slightly toward 12.5%. In the superlattice, the $T_{collapse}$ indicates a reduced dependency on the diameter. Interestingly, at 12.5% Zr and 6.5 nm the random composition displays a slightly higher $T_{collapse}$ than the superlattice. For higher Zr contents (≤15%), nanodomains can still be stabilized at 1 K but fragment into multidomain states upon heating, consistent with the progressive weakening of the ferroelectric instability under enhanced substitution.

At first glance, the bulk BZT phase diagram [27–29] suggests that increasing Zr toward ~12.5%, which raises the rhombohedral–orthorhombic (R–O) transition temperature, should monotonically stabilize the nanodomains. In practice, the nanodomain collapse temperature does not track the bulk R–O line: interfacial energetics and disorder dominate. These trends can be interpreted as follows. Average softening with Zr reduces ferroelectric stiffness and the effective domain-wall energy, driving the initial drop in $T_{collapse}$. With further substitution, disorder pinning and self-averaging counterbalance this softening, meaning quenched compositional disorder both hinders wall motion and becomes more homogeneous on mesoscopic scales, yielding the partial recovery near 12.5%. Finally, finite-size effects favor larger cylinders because the tendency of a cylindrical domain wall under tension to contract becomes weaker with increasing radius, and larger domains interact with a wider range of pinning centers. Together, these effects suppress thermally driven shrinkage and increase $T_{collapse}$. This also explains why, at large diameter, the random 12.5% composition can slightly outperform the superlattice: for extended walls, a distributed pinning landscape can provide a higher effective barrier to collapse than the coherent soft modes accessible in a periodic modulation.

Overall, pure BT offers the highest intrinsic stability. Zr substitution generally reduces it, albeit with a composition window where disorder-assisted pinning compensates. Larger diameters are systematically slightly more robust. Importantly, nontrivial topology persists over broad temperature ranges even in random BZT up to 12.5% Zr.

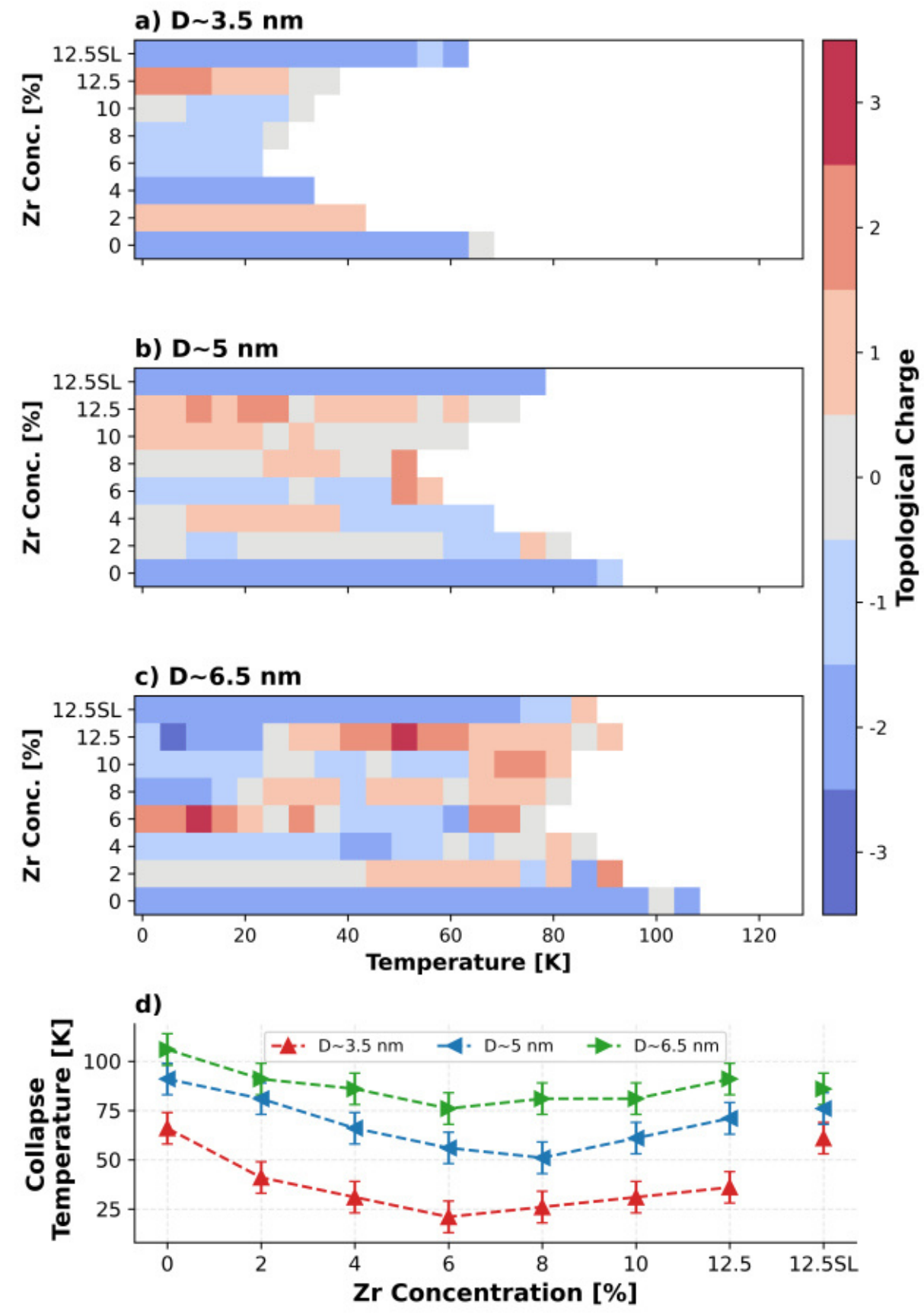

*Figure 3. Intrinsic thermal stability of nanodomains, and the associated topological textures, as a function of composition and temperature. (a) Evolution of the integrated topological charge with temperature for an induced diameter D of ~3.5 nm. (b) Same as (a) for ~5 nm. (c) Same as (a) for ~6.5 nm. (d) Collapse temperature ($T_{collapse}$) (averaged over three independent runs) as a function of Zr concentration for the three studied diameter scenarios. "12.5SL" denotes the 12.5% ordered superlattice.*

## E. Field-Stabilized Topological Textures Near Room Temperature

Finally, we ask whether ferroelectric nanodomains with nontrivial topology can be stabilized near room temperature. The previous section showed intrinsic thermal stability windows of ~50–100 K (depending on Zr content and ordering), a clear improvement over the ~1 K range of our initial studies, but still insufficient for practical applications. Our focus here is the realization and control of such textures around 300 K. To address the prerequisite, rhombohedral stability near 300 K, we first computed the finite-temperature phase diagram of the 12.5% superlattice (using the same 64×64×64 cell and Zr configuration as before), heating from 1–450 K in 1 K steps and cooling symmetrically. Figure 4a (illustrating the lattice parameters vs. temperature) shows that all four BT-like phases persist, but shifted relative to pure BT (see Refs. [27–29]). Crucially, the rhombohedral–orthorhombic transition occurs at ~320 K (cooling) and ~345 K (heating). Thus, rhombohedral order remains stable above 300 K, with modest thermal hysteresis, which fulfills the crystallographic condition for our topological textures at room temperature.

The first attempts to induce nanodomains at 293 K started from a single-domain state, where we inverted the polarization locally using a [111]-oriented cylindrical seed (as in the previous sections). Although domains formed and persisted for tens of picoseconds, they eventually relaxed back to the single-domain state due to thermal fluctuations. Under these conditions, robust zero-field retention was not achieved. As a second attempt, and following the concept of Stepková and Hlinka [14], we applied a local electric field collinear with [111] to write and stabilize a reversed cylindrical domain. A local field of 350 kV/cm reliably produced and maintained nanodomains at 293 K, for larger diameters, 250 kV/cm was sufficient, though we used 350 kV/cm (a field in the near-saturation regime of our EH–MD protocol, see Fig. S6 for the P–E loop in the Supplemental Material [39]) consistently. Figures 4b and 4c show that domains persist throughout the simulation window as long as the field remains applied (volatile operation). Around the domain circumference, nontrivial topological textures emerge, closely resembling the low-temperature patterns but now carrying a plane-integrated charge of Q = +4 at 293 K. The internal structure exhibits fragmented local charge consistent with the fractionalized motifs described earlier. Although the stabilizing field is cylindrically defined, the resulting in-plane pattern reflects the underlying rhombohedral anisotropy of the BT-based host lattice: polarization rotation and wall closure preferentially follow the symmetry-allowed $\langle 111 \rangle$ variants, yielding a threefold (3m) motif as a low-energy closure configuration. Furthermore, along the nanodomain axis, the −2/+4 alternation is absent, instead the external field enforces a uniform +4 texture. In contrast, the random 12.5% BZT composition exhibits different behavior. Although a single-domain configuration can be prepared at 293 K in the supercell, introducing the local field destabilizes it and drives a transition to a multidomain state. Thermal fluctuations together with compositional disorder are sufficiently strong to destabilize single-domain retention under bias.

As for the interpretation: in the 12.5% superlattice at 293 K, a local field E‖[111] can stabilize a single reversed cylinder, the wall adopts a non-trivial topological nature and every (111) projection yields Q=+4, indicating that $-\boldsymbol{E} \cdot \boldsymbol{P}$ overcomes thermal roughening and depolarizing costs while suppressing axial reversals. In contrast, in the 12.5% random composition, the same local field destabilizes a prepared single-domain and drives multidomain nucleation, showing that compositional disorder plus thermal fluctuations lower local nucleation barriers and favor multiple pinned walls under bias. Consequently, the random compositions do not sustain a single cylindrical domain or a uniform topological charge along $z'$ under the writing field, whereas the ordered superlattice does. Thus, at room temperature, topology is field-stabilized and volatile in the ordered case, but field-induced fragmentation dominates in the random composition, achieving robust retention in disordered samples will require raising wall stiffness and suppressing disorder-assisted nucleation, as discussed below.

Practically, this means room-temperature operation is field-programmable and volatile: domains can be written and held as long as the field is applied, and they relax when it is removed. Pathways to nonvolatile retention at 300 K include raising the domain-wall energy and suppressing thermal roughening through epitaxial strain (to stabilize the rhombohedral state and increase anisotropy), employing strain gradients, electrostatic engineering, and periodic pinning (superlattices or patterned compositional modulation) to stabilize the non-trivial topological wall structure. These routes provide concrete directions for stabilizing topological textures at room temperature without a sustaining field.

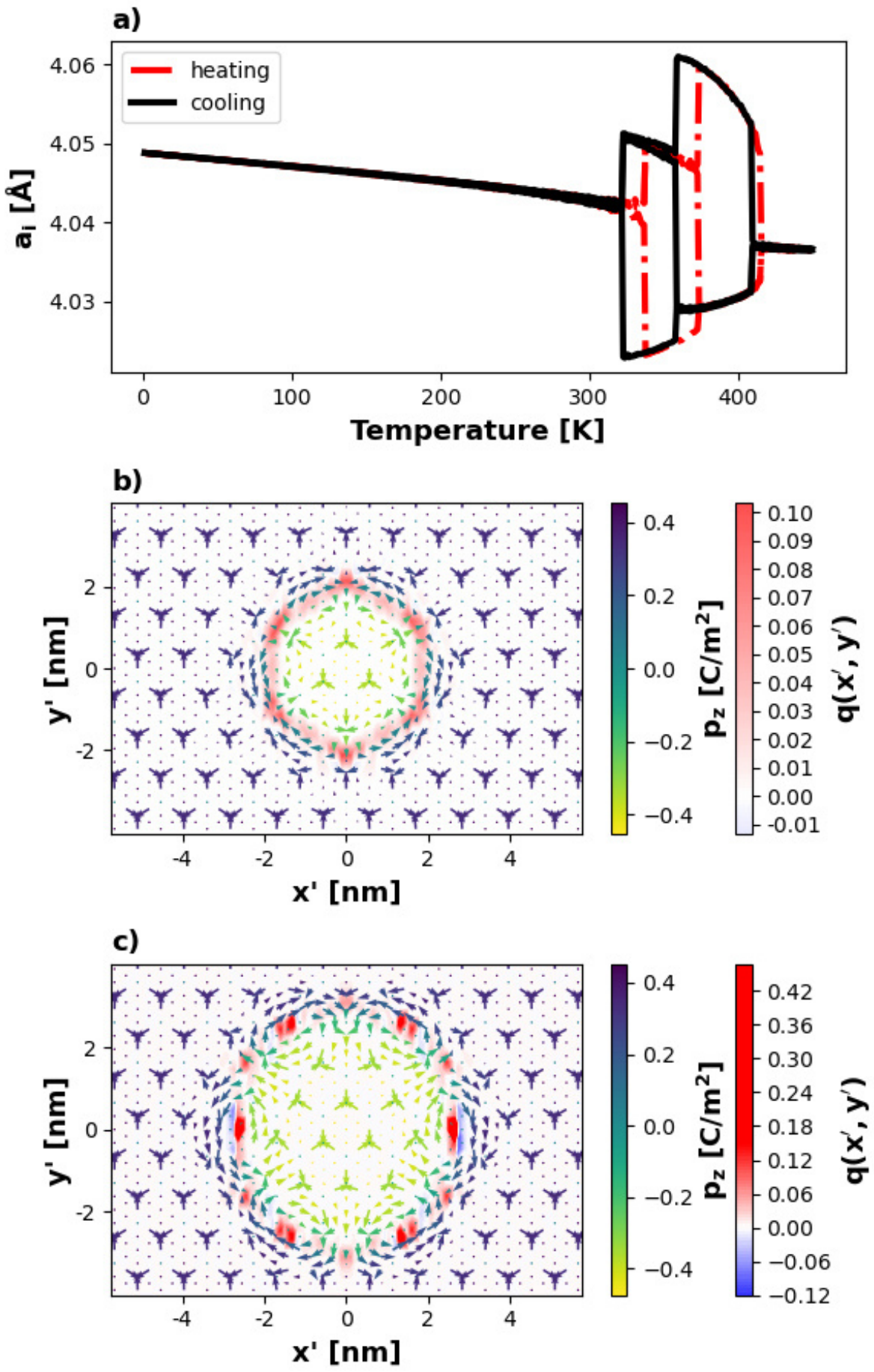

*Figure 4. Phase behavior and near room-temperature field-induced stabilization of nanodomains in the 12.5% ordered superlattice. (a) Phase diagram of the 12.5% superlattice (in terms of the lattice parameters $a_i$), showing rhombohedral stability above 300 K. (b) Skyrmion with charge Q=+4 and a diameter of ~4 nm, written and stabilized at 293 K under a local electric field of 350 kV/cm. (c) Larger Q=+4 skyrmion with a diameter of ~6 nm under the same conditions. Panels (b) and (c) use the (111) projection, with in-plane polarization vectors color-coded by the out-of-plane component and the local topological charge density overlaid. In both cases, fragmentation of the topological charge is observed.*

# IV. FINAL REMARKS AND SUMMARY

Using an EH framework benchmarked against DFT, we established that BZT supports a rich set of ferroelectric topological textures. In the 12.5% superlattice, chemical doubling along [111] yields an alternation of textures: one half hosts the canonical −2 antiskyrmion (six −1/3 topological quarks), the other a +4 skyrmion (six +2/3 topological quarks). Plane-resolved analyses show that the two chemical halves preserve the same six-vortex scaffold but exhibit distinct core-level polarization textures, yielding an integer offset of +1 per vortex in the slice charge, which is confirmed by difference maps of the local topological charge density. In random BZT, nanodomains remain inducible and stable at 1 K across 0–12.5% Zr, but disorder pins and distorts vortices, producing a skyrmion-glass-like state with spatially heterogeneous fractionalization and Q(z′) fluctuations. Thermal ramping reveals a non-monotonic intrinsic $T_{collapse}$: it drops with initial Zr substitution (minimum near 6-8%) and partially recovers toward 12.5%, reflecting competition between average ferroelectric softening and disorder pinning/self-averaging. This trend was verified across three independent random-compositions realizations, however, because finite-size effects and stochastic disorder produce configuration-dependent scatter in $T_{collapse}$, the quantitative bounds are indicative only until tested with larger supercells and a broader ensemble. Finally, near room temperature, the 12.5% superlattice is rhombohedral (cooling ~320 K, heating ~345 K). Cylindrical nanodomains cannot be retained at 293 K without bias, but a local field (~350 kV/cm, [111]) writes and holds domains whose wall selects a non-trivial topological character, giving Q=+4 on every (111) projection (no −2/+4 alternation). In contrast, the 12.5% random composition does not retain a single domain under the same bias, instead the field nucleates multidomain states due to disorder-assisted instabilities.

The unit-cell doubling in the ordered system restructures the local anisotropy landscape, enabling the integer shift of plane-resolved charge without moving vortex cores. In random compositions, quenched anisotropy and random-field components pin and shear domain walls, enabling partial charge transfers and preventing a unique texture classification. With increasing Zr, $T_{collapse}$ reflects the balance between reduced wall energy (softening) and disorder-induced pinning. At 293 K, line tension and depolarizing energy are insufficient for zero-field retention, but an external field stabilizes a non-trivial topological wall in the superlattice, whereas disorder in the composition lowers local nucleation barriers and drives fragmentation into a multi-domain state.

This work establishes BZT as a practical platform for engineered ferroelectric topology, from fractionalized (−1/3, +2/3) topological quarks at 1 K to field-stabilized Q=+4 textures near room temperature. We also introduced an optional $\mathbb{Z}_2$ bookkeeping perspective, a compact symmetry-based language for the chemically doubled motif, that captures the systematic integer (+1 per vortex) offset between chemically distinct regions. Microscopically, our data indicate that this offset is mediated by localized Bloch-point–like conversion events, enabled by Zr-assisted suppression of the polarization amplitude ($|\boldsymbol{p}|\to 0$) at/near vortex axes within the interfacial region. What remains to be clarified is the detailed spatial and temporal statistics of these conversion events, how localized they remain as a function of temperature, disorder, and skyrmion diameter, and how the availability of multiple low-$|\boldsymbol{p}|$ sites in broader cores affects the robustness of the alternation.

On the materials side, the proposed 12.5% ordered superlattice provides a clean phenomenology but may be difficult to synthesize with long-range chemical order across device-relevant areas. Fortunately, several alternative routes could possibly stabilize the rhombohedral phase near 300 K without strict chemical ordering: epitaxial biaxial strain [54] (via substrate choice and orientation), compositional tuning or gradients [27,35] (chemical pressure), and electrostatic engineering [55] (improved screening to suppress depolarizing fields). These approaches could also increase domain-wall energy and can enhance retention.

It is worth noting the broader potential of the topological textures discussed here and in the literature. If the internal fractional structure could be externally addressed and read out, topology would move beyond a binary scheme toward true multi-state functionality. Instead of simply switching a nanodomain on or off, information could also be stored in its internal configuration, an appealing prospect for memory concepts as well as neuromorphic schemes. Realizing this remains a longer-term vision: experimental confirmation of such textures and strategies for their stabilization are the immediate priorities, with epitaxial strain, chemical substitution, and related approaches offering promising routes.

In summary, BZT emerges as a versatile playground for ferroelectric topology, where both ordered and random chemistries support textures with rich fractional internal structure and clear thermal stability windows from cryogenic temperatures to near room temperature, including field stabilized states at 293 K. These results identify strain, electrostatics, and compositional patterning as promising avenues for systematic studies of the creation, manipulation, and readout of topological textures, setting the stage for broader exploration.

# ACKNOWLEDGMENTS

The author gratefully acknowledges Jiří Hlinka (Czech Academy of Sciences) and Maxim N. Popov (Materials Center Leoben) for valuable discussions. The author gratefully acknowledges the financial support under the scope of the COMET program within the K2 Center “Integrated Computational Material, Process and Product Engineering (IC-MPPE)” (Project 886385). This program is supported by the Austrian Federal Ministries for 718 Climate Action, Environment, Energy, Mobility, Innovation, and Technology (BMK) and for Digital and Economic Affairs (BMDW), represented by the Austrian research funding association (FFG), and the federal states of Styria, Upper Austria, and Tyrol.

# Supplemental Material

## Topological Textures in Zr-Substituted Barium Titanate

Florian Mayer[1*]

[1]Materials Center Leoben Forschung GmbH, Vordernberger Strasse 12, 8700 Leoben, Austria

*florian.mayer@mcl.at

# I. APPLIED EFFECTIVE HAMILTONIAN

The basis for the molecular dynamics (MD) simulations of the Zr-substituted BT ($BaTiO_3$) systems is an effective Hamiltonian in the local-mode framework. Each 5-atom unit cell at lattice vector $\boldsymbol{R}$ carries (i) a local soft mode $\boldsymbol{u}(\boldsymbol{R})$ (proportional to the local dipole), and (ii) a set of acoustic displacement variables $\boldsymbol{w}(\boldsymbol{R})$ that represent inhomogeneous strain/translation degrees of freedom. In addition, the simulation cell has homogeneous strain variables $\eta_i$ (i=1…6, Voigt notation). For substituted systems, a discrete species label $\sigma$ is assigned to each B-site unit cell (e.g., $\sigma$ = Ti or Zr).

The baseline Hamiltonian used for the parent compound BT follows the Nishimatsu-type [1,2] EH form (with the parameterization described in Ref. [3] and associated references therein). It is written as

$$
\begin{aligned}
H_{BTO}^{eff} = \frac{M_{dipole}^{*}}{2}\sum_{R,\alpha}\dot{u}_{\alpha}^{2}(\boldsymbol{R}) \\
+ \frac{M_{acoustic}^{*}}{2}\sum_{R,\alpha}\dot{w}_{\alpha}^{2}(\boldsymbol{R}) + V^{self}(\{\boldsymbol{u}\}) + V^{dpl}(\{\boldsymbol{u}\}) + V^{short}(\{\boldsymbol{u}\}) \\
+ V^{elas,homo}(\eta_1,\dots,\eta_6) + V^{elas,inho}(\{\boldsymbol{w}\}) + V^{coupl,homo}(\{\boldsymbol{u}\},\eta_1,\dots,\eta_6) \\
+ V^{coupl,inho}(\{\boldsymbol{u}\},\{\boldsymbol{w}\}) - Z^{*}\sum_{R}\epsilon\cdot u(\boldsymbol{R})
\end{aligned}
\tag{S1}
$$

Here, $M_{dipole}^{*}$ and $M_{acoustic}^{*}$ are effective masses for the local-mode and acoustic variables, $Z^{*}$ is the Born effective charge associated with the local mode, and $\epsilon$ is the applied electric field. The energy terms $V^{self}$, $V^{dpl}$, $V^{short}$, $V^{elas,homo}$, $V^{elas,inho}$, $V^{coupl,homo}$, and $V^{coupl,inho}$ have the standard EH meanings [1–4] (local-mode self-energy, dipole–dipole interaction, short-range interaction, elastic energies, and strain–mode couplings, respectively).

To model substitution on the B-site, we treat dopants as a perturbation of the parent BT Hamiltonian. The total Hamiltonian is

$$
H_{total}^{eff}(\{\boldsymbol{u}\},\{\boldsymbol{w}\},\eta_i,\{\sigma\}) = H_{BTO}^{eff}(\{\boldsymbol{u}\},\{\boldsymbol{w}\},\eta_i) + H^{peturb.}(\{\boldsymbol{u}\},\{\boldsymbol{w}\},\{\sigma\})
\tag{S2}
$$

with the perturbation decomposed as

$$
H^{peturb.} = \Delta T(\{\boldsymbol{u}\},\{\boldsymbol{w}\},\{\sigma\}) + \Delta V^{self}(\{\boldsymbol{u}\},\{\sigma\}) + \Delta V^{dpl}(\{\boldsymbol{u}\},\{\sigma\}) + V^{aux}(\{\boldsymbol{u}\},\{\boldsymbol{w}\},\{\sigma\})
\tag{S3}
$$

Here, $\Delta T$ accounts for species-dependent effective masses (kinetic terms), $\Delta V^{self}$ modifies the local-mode self-energy on substituted sites, $\Delta V^{dpl}$ accounts for species-dependent Born effective charges in the long-range dipolar interaction, and $V^{aux}$ is an auxiliary "spring" correction that captures local chemistry-induced interaction changes beyond the parent Hamiltonian. All corrections are described in detail in Ref. [3]. The explicit polynomial form used for the species-dependent self-energy correction can also be found in Ref. [3].

Chemistry-induced modifications of interactions with neighboring cells are modeled by the first-order auxiliary term

$$
V^{aux}(\{\boldsymbol{u}\},\{\boldsymbol{w}\},\sigma) = \sum_{R}\sum_{T} Q_{\boldsymbol{T},\boldsymbol{R}}(\sigma_{\boldsymbol{T}})\boldsymbol{e}_{\boldsymbol{T},\boldsymbol{R}}\cdot\boldsymbol{u}(\boldsymbol{R}) + \sum_{R}\sum_{T} S_{\boldsymbol{T},\boldsymbol{R}}(\sigma_{\boldsymbol{T}})\boldsymbol{f}_{\boldsymbol{T},\boldsymbol{R}}\cdot\boldsymbol{w}(\boldsymbol{R})
\tag{S4}
$$

The index $\boldsymbol{T}$ runs over neighbors up to the third-nearest-neighbor shell (3NN), $\boldsymbol{e}_{T,R}$ is the unit vector joining site $\boldsymbol{T}$ to the center of $\boldsymbol{u}(\boldsymbol{R})$, and analogously $\boldsymbol{f}_{T,R}$ joins $\boldsymbol{T}$ to the center of $\boldsymbol{w}(\boldsymbol{R})$. The coefficients $Q_{\boldsymbol{T},\boldsymbol{R}}$ and $S_{\boldsymbol{T},\boldsymbol{R}}$ depend on the species at $\boldsymbol{T}$ and are obtained by first-principles fitting as described in Ref. [3].

To incorporate dopant-induced lattice expansion/contraction at the homogeneous-strain level, the model applies a composition-dependent hydrostatic pressure correction P(x) (with x the dopant concentration). For BZT, Ref. [3] reports a linear correction

$$P(x) = -19.5x\ [GPa] \tag{S5}$$

derived from DFT trends of lattice parameter vs pressure (pure BT) combined with lattice parameter vs concentration (BZT). In practice, the homogeneous strain $\eta_i$ is therefore allowed to relax self-consistently under this imposed hydrostatic pressure, rather than under strictly zero external stress.

All simulations in the present work are performed with three-dimensional periodic boundary conditions. The order parameter used for the topological characterization is the local mode (or equivalently the local polarization) field; in the main text we construct the unit-vector field $\boldsymbol{n}(\boldsymbol{R})$ by normalizing the local mode/polarization, $\boldsymbol{n}(\boldsymbol{R}) = \boldsymbol{u}(\boldsymbol{R})/|\boldsymbol{u}(\boldsymbol{R})|$ (or $\boldsymbol{n}(\boldsymbol{R}) = \boldsymbol{p}(\boldsymbol{R})/|\boldsymbol{p}(\boldsymbol{R})|$, with $\boldsymbol{p}(\boldsymbol{R}) = Z^*(\sigma_{\boldsymbol{R}})\boldsymbol{u}(\boldsymbol{R})/\Omega_0$ and $\Omega_0$ being the unit-cell volume), and compute slice-resolved topological charges using a lattice discretization consistent with this EH description.

# II. THE 12.5% BZT SUPERLATTICE

The basis of the first study in the main paper is a superlattice created by introducing Zr into BT. The $ABO_3$ perovskite structure of BT is well known and will not be discussed in further detail here. The symmetry of the superlattice is analyzed using the cubic phase as a reference, since the primary focus of subsequent investigations is the ordering of the B-sites. This higher-level chemical ordering remains unchanged even in the studied rhombohedral phase. The superlattice considered here is conceptually simple. We begin with pure BT, double the unit cell to 2×2×2, and substitute one B-site atom with Zr. This results in an exact Zr concentration of 12.5%. Consequently, translational symmetry is broken while the point symmetry is preserved, resulting in a doubling of the unit cell. In simple terms, every second B-site along the Cartesian directions is now occupied by Zr instead of Ti (see Figure S1 for an illustration of the B-site ordering). Of particular interest is the [111] crystallographic direction. Along this direction, the unit cell is also effectively doubled, and defines a superlattice translation vector along $\langle 111 \rangle$ as: $\boldsymbol{c}_{[111]} = 2a\sqrt{3}\hat{\boldsymbol{e}}_{111}$, where $\hat{\boldsymbol{e}}_{111}$ denotes the unit vector in the $\langle 111 \rangle$ direction and $a$ is the lattice constant of pure BT. However, care must be taken: the smallest repeatable geometric unit along [111] consists of three (111) planes, similar to the ABC stacking found in face-centered cubic (fcc) structures. The chemical motif doubles this sequence, yielding a Ti–Zr–Ti–Zr–Ti–Zr stacking. In other words, every second (111) plane contains both Ti and Zr ions, while the others remain purely Ti.

Despite this chemical modulation, the new unit cell along the [111] direction can be divided into two distinct halves based on geometric lattice sites. The first half exhibits a Ti–Zr–Ti stacking, and the second half a Zr–Ti–Zr stacking. Figure S1c and S1d highlight these regions. Both segments share identical geometric B-site positions, however, the sequence of Zr/Ti-containing planes is reversed. These two chemically distinct, yet geometrically equivalent, halves are of critical importance for the investigation of topological textures presented in the main paper and can thus be further characterized. As demonstrated in previous studies [5], the three geometrically distinct (111) planes constitute the fundamental unit capable of hosting topological structures. In principle, a projection of these planes is used to compute the Pontryagin topological density. Although the chemical motif doubles the unit cell, the geometric lattice positions in both halves remain unchanged. This means that projecting all six planes in two dimensions would lead to overlapping lattice sites, which is undesirable for our topological analysis. Therefore, the original geometric unit remains the relevant structural element for our study of topological textures, at least when we restrict the study to the 2D case, while the chemical ordering is considered a background that modifies local anisotropy.

To investigate the applicability of our studies on the chemical superlattice, we first performed DFT calculations using a 2×2×2 supercell corresponding to the new unit cell of the superlattice, allowing for full structural relaxation. The key question is how the Zr ions influence the pristine BT easy axes. To address this, we analyzed the fully relaxed DFT structure and approximated the local polarization by evaluating the displacement of B-site cations relative to the surrounding oxygen octahedra. The results are shown in Figures S1a and S1c, where the unit cell has been doubled (to a 4×4×4 cell) for visualization purposes. These reveal that Zr is only weakly polar, exhibiting a significantly reduced off-centering compared to Ti. However, Zr exerts a notable influence on the surrounding Ti-centered unit cells, distorting their easy axes. In general terms, Zr perturbs the local polarization landscape by altering the preferred directions of polarization in neighboring Ti cells. This effect has been extensively studied in prior work [3,6], to which we refer here for further details. For our purposes, the critical observation is that Zr induces a new local anisotropy while maintaining the threefold rotational symmetry and an overall net polarization along [111].

For the sake of comparison and to set the stage for the study of topological structures using MD within the EH framework, we performed a benchmark simulation. A 64×64×64 supercell with the periodic Zr distribution was thermalized at 1 K for 800 ps, followed by a 200 ps production run using the MD settings from Section II. The resulting polarization field was analyzed, and a representative 4×4×4 region is shown in Figures S1b and S1d. For direct comparison with DFT, polarization vectors from both methods were normalized to their respective maxima. The EH model closely matches DFT results, reproducing both the suppression of polarization at Zr sites and the in-plane reorientation of nearby Ti dipoles. This confirms the EH model's accuracy in capturing the ground-state behavior of the BZT12.5 superlattice and its suitability for exploring topological phenomena.

All density functional theory (DFT) calculations were carried out using the VASP [7–11] simulation package and employing the PBEsol [12] exchange-correlation functional. Projector-augmented wave (PAW) potentials [7] were used with the following valence electronic configurations: Ba ($5s^2$ $5p^6$ $6s^2$, 10 valence electrons), Ti ($3s^2$ $3p^6$ $4s^2$ $3d^2$, 12 valence electrons), Zr ($4s^2$ $4p^6$ $5s^2$ $4d^2$, 12 valence electrons), and O ($2s^2$ $2p^4$, 6 valence electrons). All calculations were performed on 2×2×2 supercells using a 4×4×4 Monkhorst–Pack k-point mesh. The plane-wave energy cutoff was set to 620 eV, and both the k-grid and cutoff energy were validated through convergence testing. Electronic self-consistency was achieved with an energy convergence threshold (EDIFF) of 1E-8 eV. Ionic relaxations were performed using the conjugate-gradient algorithm (IBRION = 2), allowing for full relaxation of both ionic positions and lattice parameters (ISIF = 3). Structural optimizations were considered converged when the maximum residual force fell below 1E-4 eV/Å and the stress tensor components were reduced to below 1E-2 MPa.

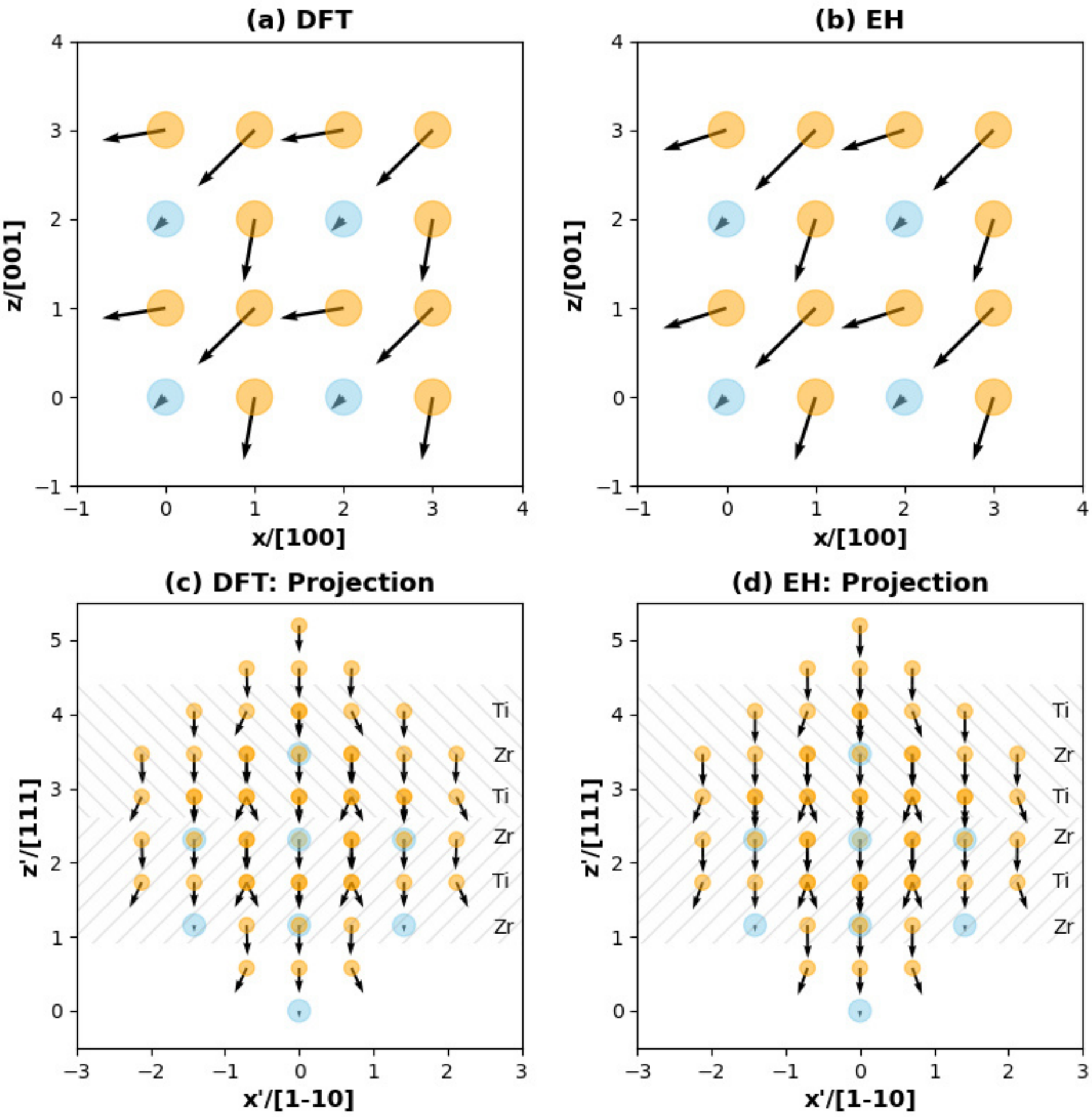

*Figure S1. Visualization of the B-site cations in the 12.5% Zr-substituted BT superlattice. Blue spheres represent Zr ions, yellow spheres represent Ti ions. Arrows indicate normalized polarization vectors centered at the B-sites, scaled relative to the maximum polarization magnitude. Panels a and b show the x–z plane: (a) with polarization vectors from DFT calculations, and (b) from EH simulations. Panels c and d present projections: (c) with DFT-derived polarization, and (d) with EH-derived polarization. The hatched areas indicate the two regions with different Zr distributions but identical geometric planes, which together form the new doubled unit cell.*

# III. ADDITIONAL VISUALIZATION OF |P| SUPPRESSION AND FRACTIONAL-CHARGE VORTEX TEXTURES

This section provides additional microscopic evidence supporting the discussion in Sections III.A and III.B of the main text. In particular, the observed alternation between the plane-integrated slice charges −2 and +4 (and the corresponding redistribution into six −1/3 versus six +2/3 topological quarks) implies that the plane-resolved topological charge must change by an integer across the chemical interface. Such an integer change cannot occur continuously for a strictly normalized order-parameter field $\boldsymbol{n}(\boldsymbol{R})$, it requires a local breakdown of the normalization, i.e., a singular conversion event in which the polarization amplitude becomes strongly suppressed. In Section III.B we therefore argued that the charge transfer is naturally mediated by Bloch-point–like polarization singularities (local $|\boldsymbol{p}|\to 0$) within the interfacial conversion region.

Figure S2 replots the same cross sections as Figures 1c and 1d of the main text, now with the polarization magnitude $|\boldsymbol{p}|$ shown as an interpolated background colormap. The interpolation is used solely to visualize the spatial distribution of $|\boldsymbol{p}|$ beyond discrete lattice-site values, the arrow field continues to represent the local polarization vectors (with arrow color encoding the out-of-plane component).

A key observation is that Zr-substituted unit cells (marked by circles) frequently coincide with pronounced local suppression of $|\boldsymbol{p}|$. In the conversion region, the suppression at/near the vortex axis is typically much stronger than the background reduction at generic Zr-centered sites away from the cores, which allows the near-nodes associated with charge transfer to be distinguished from chemically soft but topologically inactive low-$|\boldsymbol{p}|$ locations. This reflects the chemically "soft" polar response of Zr-centered environments in BT-based systems, which can support configurations with very small local polarization amplitude. Importantly, in the +4 half (Figure S2a) the strongest near-nodes occur at/near vortex axes within the interfacial region, providing precisely the type of localized $|\boldsymbol{p}|\rightarrow 0$ events required for an integer change in the plane-resolved topological charge. In contrast, while regions of reduced |p| also occur in the −2 half (Figure S2b), again typically associated with Zr sites, they do not coincide with the vortex-axis conversion sites in the same way in this cross section. This is consistent with the main-text picture that the −2 antiskyrmion texture remains closer to the BT baseline, whereas the +4 half accesses an interfacial conversion pathway enabled by chemically available low-$|\boldsymbol{p}|$ sites.

This comparison also clarifies an important point: low $|\boldsymbol{p}|$ is necessary but not sufficient for topological charge transfer. Zr substitution may create many low $|\boldsymbol{p}|$ locations, but only those that lie on (or sufficiently near) the relevant vortex axis within the conversion region participate in the source/sink behavior of the topological density.

Finally, these observations are fully compatible with the role of Section III.B's $\mathbb{Z}_2$ construction. The $\mathbb{Z}_2$ "phase convention" is used as optional bookkeeping that partitions the two chemical halves into inequivalent sectors, it does not itself provide a microscopic mechanism for changing $Q(z')$. The microscopic conversion is instead realized by the localized suppression of $|\boldsymbol{p}|$ highlighted in Figure S2, which renders $\boldsymbol{n}(\boldsymbol{R})$ ill-defined and permits an integer jump in the plane-resolved charge.

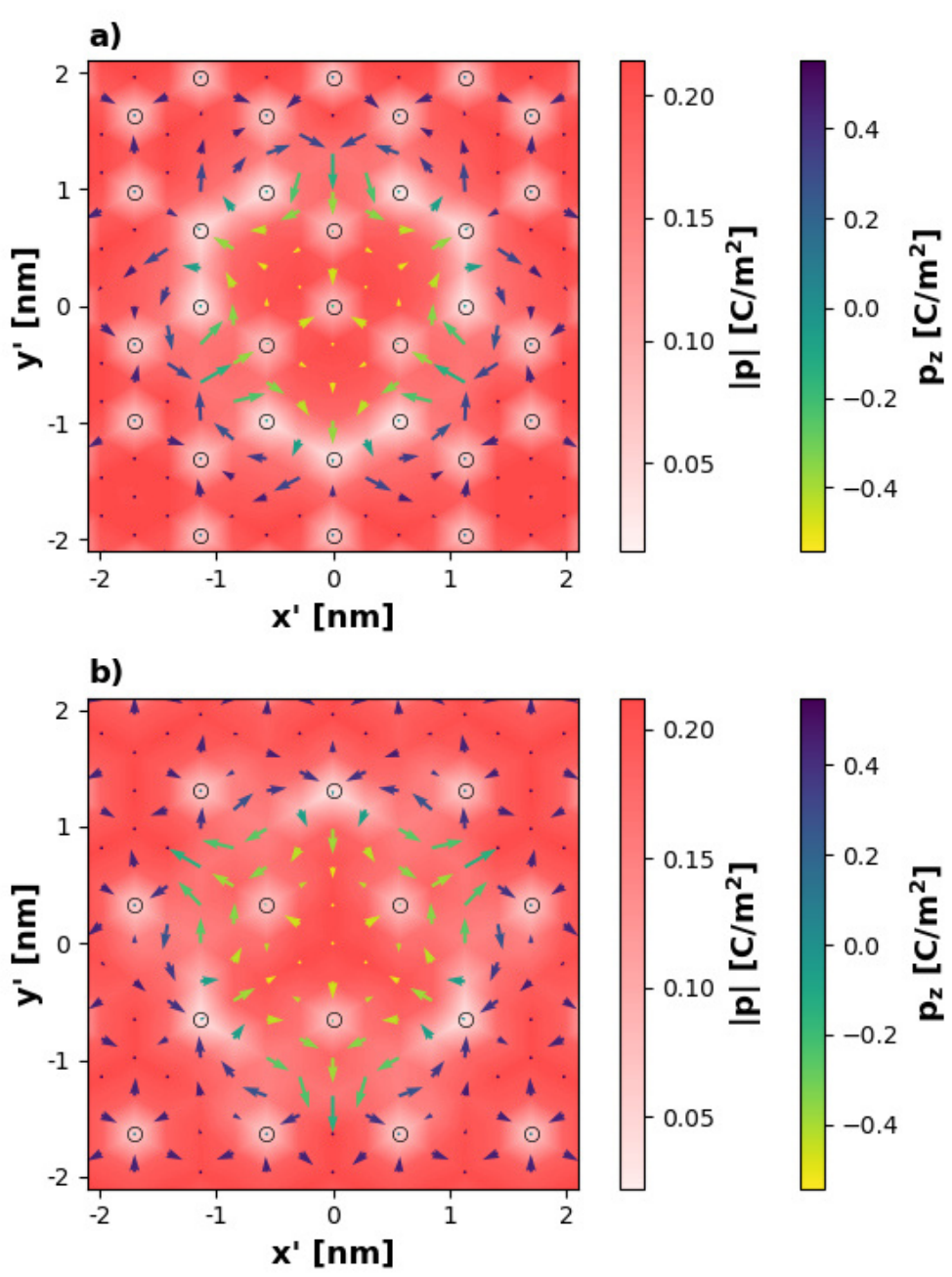

*Figure S2. Zr-assisted suppression of the polarization magnitude at vortex cores. (a) In the +4 half, several vortex cores intersect Zr-centered unit cells (circles). At these sites the polarization amplitude $|\boldsymbol{p}|$ is strongly suppressed, providing chemically "soft" locations where the order parameter can approach $|\boldsymbol{p}|\rightarrow 0$ and thereby facilitate an integer shift of the plane-resolved topological charge. (b) In the −2 half, the Zr-centered cells do not coincide with the vortex cores in this cross section, and correspondingly no comparable near-vanishing $|\boldsymbol{p}|$ is observed at the core positions. Circles mark Zr-centered cells. The panels show the same cross sections as in Figure 1. Arrows indicate the local polarization vectors; arrow color encodes the out-of-plane component, while the interpolated background color map encodes the polarization magnitude $|\boldsymbol{p}|$.*

Figure S3 provides a zoomed, core-resolved visualization of representative vortices associated with +2/3 (Figure S3a) and −1/3 (Figure S3b) quarks, taken from the regions shown in Figures 1c and 1d of the main text. These panels emphasize that, while the two halves share the same global six-vortex scaffold (threefold motif), the core-level polarization textures differ significantly. In particular, the +2/3 vortex example exhibits a strongly suppressed (near-vanishing) $|\boldsymbol{p}|$ at the core, whereas the −1/3 example does not show a comparably suppressed core in the corresponding cross section. Circles again indicate Zr-substituted unit cells, highlighting the chemical context in which near-nodes are most readily realized.

Together, Figures S2 and S3 therefore support two main conclusions that directly reinforce the main paper: (i) the integer (+1 per vortex) offset between the −2 and +4 halves is consistent with localized Bloch-point–like conversion events enabled by Zr-mediated suppression of the polarization amplitude, and (ii) despite the shared 3m scaffold at large scales, the two textures are not identical, they differ in their core-level structure and in the distribution of skyrmion density that underlies the fractional −1/3 versus +2/3 hotspot weights.

As noted in Section III.B, this scaffold-based organizing picture is expected to weaken when the vortices broaden (e.g., at larger skyrmion diameters or under strong disorder), because a larger core samples more Zr-centered cells and thus offers multiple low $|\boldsymbol{p}|$ candidate sites. In that regime the conversion can become less localized along z′, and the clean −2/+4 alternation is correspondingly blurred, making a purely real-space description more appropriate.

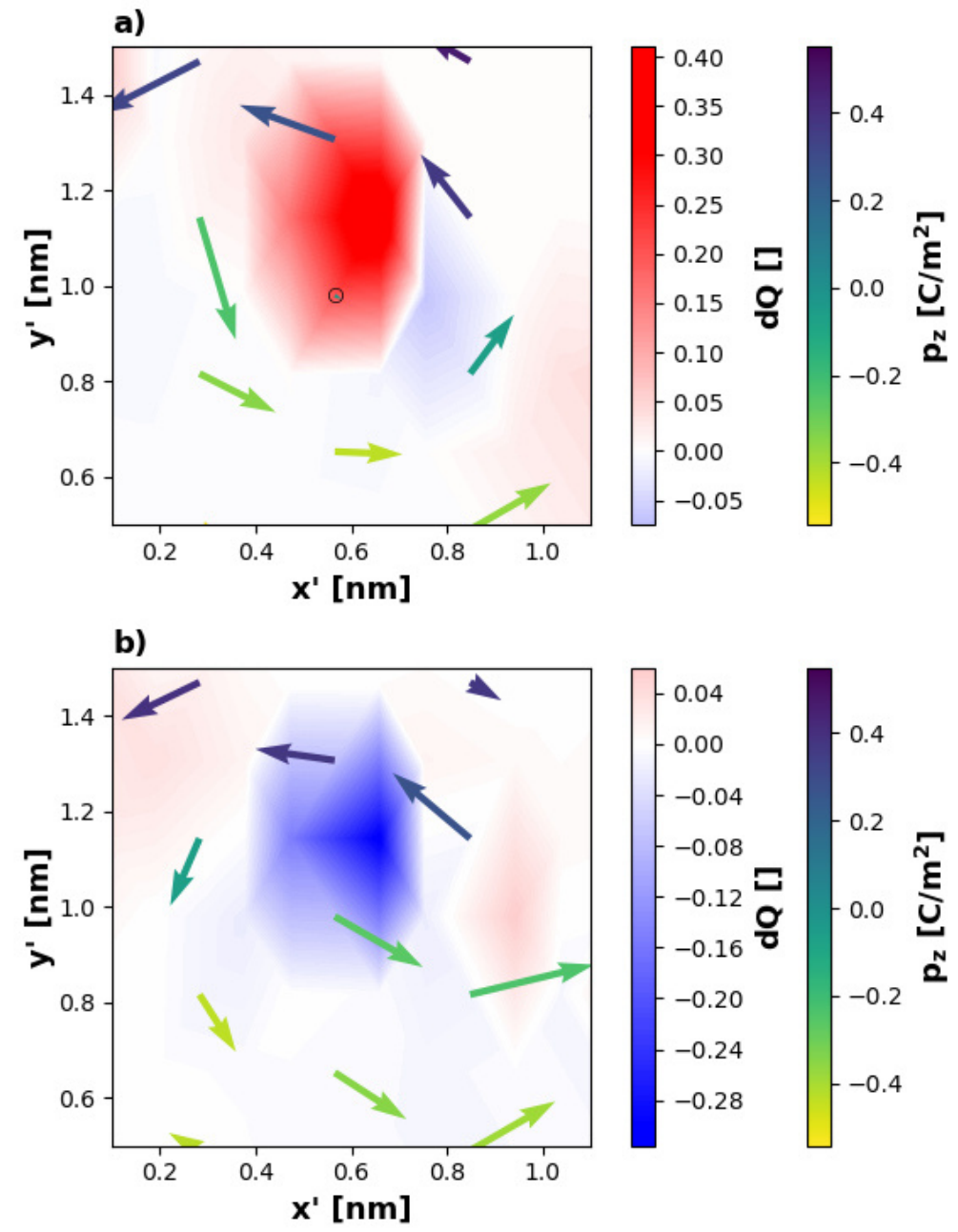

*Figure S3. Core-resolved comparison of fractional-charge vortices in the +4 and −2 halves. (a) Zoom-in of a representative vortex associated with a +2/3 quark in the +4 half, exhibiting a strongly suppressed (near-vanishing) polarization magnitude at the core. (b) Zoom-in of a representative vortex associated with a −1/3 quark in the −2 half, which does not display a comparably suppressed core in this cross section. These panels are magnified views of the regions shown in Figs. 1c and 1d of the main text and highlight that, despite sharing the same threefold six-vortex scaffold at larger scales, the core-level polarization textures differ between halves. Circles mark Zr-substituted unit cells.*

# IV. TOPOLOGICAL CHARGE DIFFERENCE

To complement the real-space mechanism discussed in Section III.B of the main text, we quantify directly whether the two chemically distinct halves (Ti–Zr–Ti vs Zr–Ti–Zr) differ by an integer offset of +1 per vortex in the plane-integrated (slice) topological charge, while preserving the same six-core scaffold in the plane. This test is purely diagnostic: it does not assume a particular microscopic route for the charge transfer, but it does require that corresponding vortex cores can be paired across the interface (i.e., remain sufficiently co-located in (x′,y′)) so that the comparison is meaningful. We evaluate the local plane-resolved topological charge density q(x′,y′) on representative (111) planes associated with the upper (+4) and lower (−2) chemical halves, and form the difference map

$$dq(x', y') = q_U(x', y') - q_L(x', y').$$

(S6)

Here $q_U$ and $q_L$ denote the densities computed on the planes in the +4 and −2 halves, respectively.

The resulting $dq(x', y')$ (Figure S4, middle panel) exhibits six sharp peaks located at the vortex-core positions, while $dq$ is near zero elsewhere. Integrating $dq$ over small disks centered at each peak yields approximately +1 per vortex within numerical precision. This demonstrates that the change of the plane-integrated charge between the two chemical halves is highly localized to the six vortex cores and is consistent with an integer +1 transfer per core, i.e. the mapping −1/3→+2/3 at corresponding scaffold positions.

This observation supports the main-text statement that the two halves can be viewed (optionally) as belonging to inequivalent sectors under the chemically imposed $\mathbb{Z}_2$ sign structure (a π reference-frame flip), providing compact bookkeeping for the systematic −2→+4 alternation. Importantly, the difference-map test establishes the integer offset but does not, by itself, specify the microscopic conversion mechanism along z′. In the main text and in the preceding section we present evidence that the conversion is accompanied by strong local suppression of the polarization amplitude at/near vortex axes within the interfacial region, consistent with Bloch-point–like singular conversion events (local $|\boldsymbol{p}|\to 0$) enabling the required jump of the normalized field $\boldsymbol{n}(\boldsymbol{R})$.

Finally, we note that the analysis presumes that corresponding cores remain sufficiently co-located so that the six peaks in dq(x′,y′) can be cleanly identified and integrated. Strong disorder, pronounced lateral wandering, or significantly larger diameters (where the core broadens and additional low-$|\boldsymbol{p}|$ sites become available) can smear the peaks and blur the −2/+4 alternation, in which case a purely real-space description becomes more appropriate, as discussed in Section III.B of the main text.

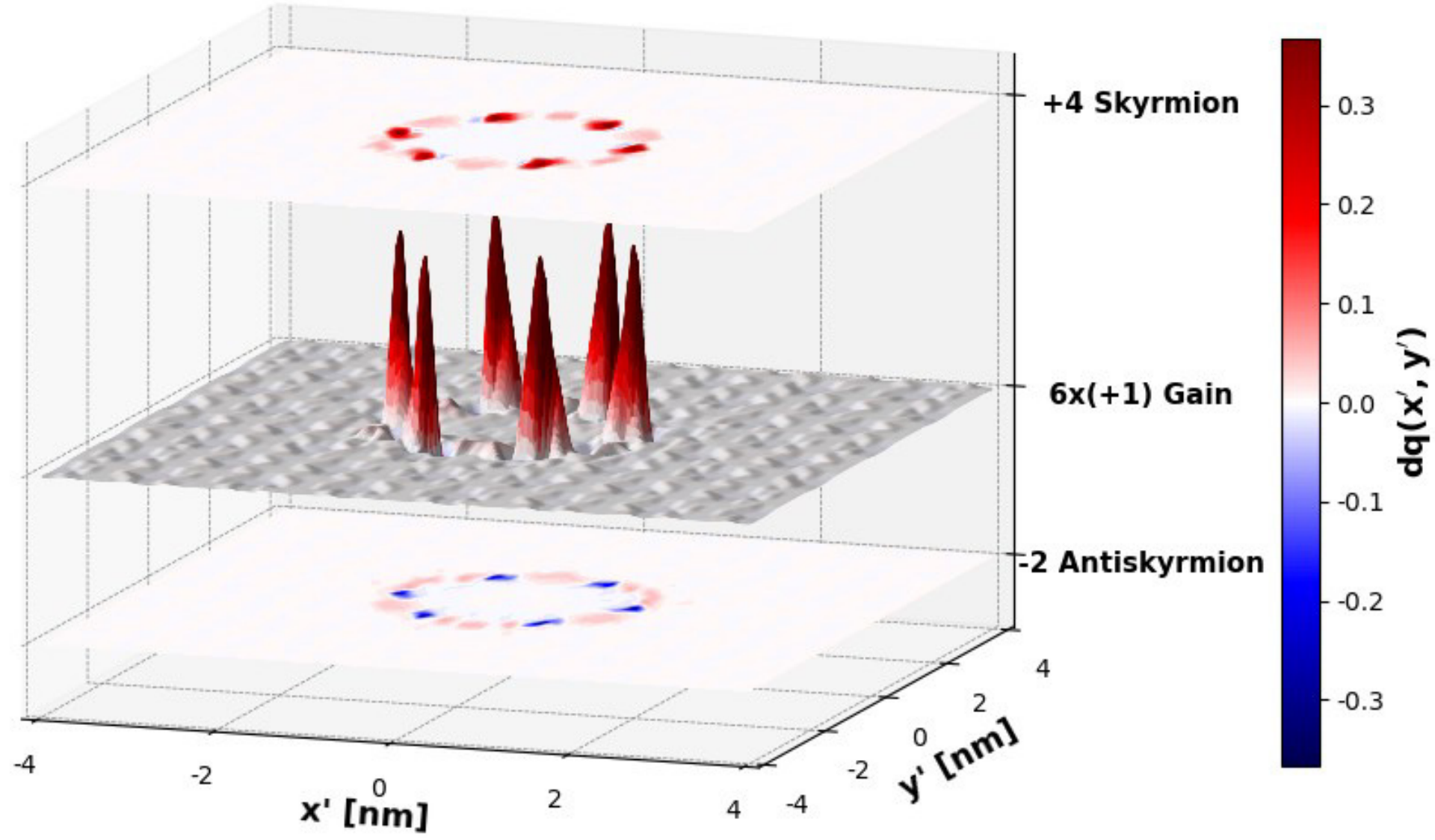

*Figure S4. Demonstration of topological charge transfer across the chemically induced interface. The bottom layer shows the local topological charge density of a –2 antiskyrmion, composed of six topological quarks, each carrying a charge of –1/3. The top layer displays a +4 skyrmion, similarly fragmented into six quarks, now each carrying +2/3 charge. The middle layer presents a surface plot of the difference between the topological charge densities, highlighting six distinct peaks, each contributing a net charge of +1. This confirms that each quark gains exactly one unit of charge upon crossing the chemical interface consistent with the optional $\mathbb{Z}_2$ sign-structure bookkeeping introduced in Section III.B. Note: All integrations are performed on the unsmoothed lattice data, smoothing is used only for visualization. In the underlying lattice data, the six difference peaks are co-located with the vortex-core scaffold to within the lattice resolution.*

# V. RANDOM CASE: EVOLUTION ALONG NANODOMAIN AXIS

This section examines the axial evolution of the topological structure (diameter ~4 nm), complementing the discussion in Section III.C of the main text. For each composition, we compute the plane-resolved topological charge by slicing the simulation cell with (111) planes normal to the nanodomain axis, $z' \parallel [111]$. At each slice position $z'$, we calculate the local topological density and integrate over the cross-section to obtain the total topological charge. The slicing plane is then shifted along $z'$ in layer increments, and the procedure is repeated, allowing us to track topological charge $Q$ along the core. All results are averaged over the final 200 ps of each 1 K trajectory. In pure BT (Figure S5a), the $-2$ antiskyrmion is axially uniform, means $Q(z')$ remains constant at $-2$, and the six-vortex scaffold shows no axial rearrangement over the analyzed core segment. In random BZT with 2% Zr (Figure S5b), axial fluctuations emerge. Local anisotropy variations induced by Zr distort vortex cores and create narrow interfacial sheets where q changes abruptly. Consequently, $Q(z')$ exhibits discrete integer steps, typically between +1 and $-2$, while the six-vortex scaffold persists. This behavior resembles a stacked skyrmion-glass-like state: an axially coupled but aperiodic sequence of plane textures. At higher substitution levels, 6% and 12.5% Zr (Figures S5c, S5d), defect fields pin and shear vortex lines more strongly. Both the amplitude and frequency of axial variations increase, and $Q(z')$ develops plateaus drawn primarily in the range from -3 to +2. In this regime, the textures can no longer be described by a single $Q$ value along the axis. Instead, they form an axially heterogeneous topological state. These results substantiate the main-text claims: (a) pure BT displays constant plane-resolved charge along the core at 1 K, (b) random Zr substitutions with moderate concentrations induce composition-dependent axial fluctuations of $Q$ while preserving the six-vortex scaffold, and (c) the random case is best described as a skyrmion-glass-like stack of plane textures whose local topological charge is set by the nearby defect landscape.

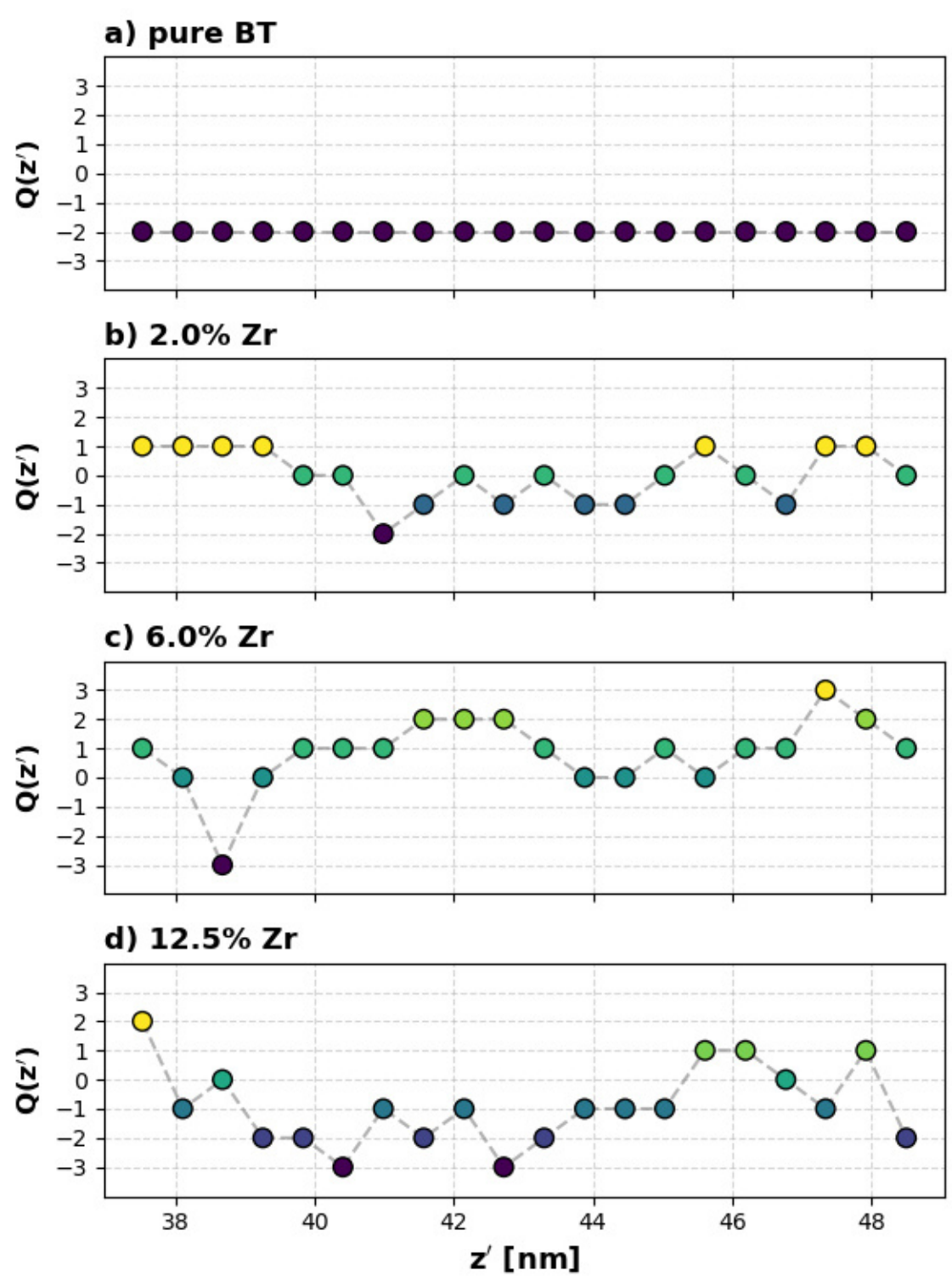

*Figure S5. Evolution of topological charge along a selected region of the nanodomain axis z' for BZT compositions with different Zr concentrations: (a) pure BT, (b) 2% randomly distributed Zr, (c) 6% randomly distributed Zr, and (d) 12.5% randomly distributed Zr.*

# VI. HYSTERESIS OF 12.5% BZT SUPERLATTICE

To clarify how the local write fields used in Sec. III.E compare to the intrinsic switching behavior of the ordered 12.5% BZT superlattice, we computed a polarization–electric-field (P–E) loop at T=293 K using the same effective-Hamiltonian MD protocol and supercell size as in the main text. The electric field was applied along [111]. For computational tractability, the loop in Figure S6 was obtained by driving the system with a time-dependent field of peak amplitude $E_{max}$=800 kV/cm at a frequency of 1 GHz (i.e., a dynamic hysteresis measurement within the EH–MD timescale).

Figure S6 shows a well-defined hysteresis loop with a saturation polarization of approximately $P_s$≈0.25 C/m$^2$, where P denotes the supercell-averaged polarization projected onto the field direction (equivalently, the average of the local dipoles along [111] divided by the unit-cell volume). The saturation value is reduced compared with pure BT, consistent with the weaker local polarity of Zr-centered cells within the BT matrix.

From the same loop we estimate a dynamic coercive field of $E_c$≈170 kV/cm (defined as the field magnitude at which the projected polarization crosses zero during the driven cycle). We emphasize that this coercive field is protocol-dependent (frequency and timescale) and therefore serves as an order-of-magnitude reference for the fields employed in Section III.E. In particular, the local write field $E$=350 kV/cm used to stabilize the cylindrical nanodomains lies well above $E_c$ and already in the near-saturation regime of the loop. This explains why, under sustained bias, the polarization magnitude inside the field-stabilized region and in the surrounding rhombohedral matrix are comparable in absolute value (with opposite sign), as observed in Figure 4 of the main text: the applied field is sufficiently strong to overcome thermal roughening and depolarizing costs while keeping the polarization close to its saturated magnitude along [111].

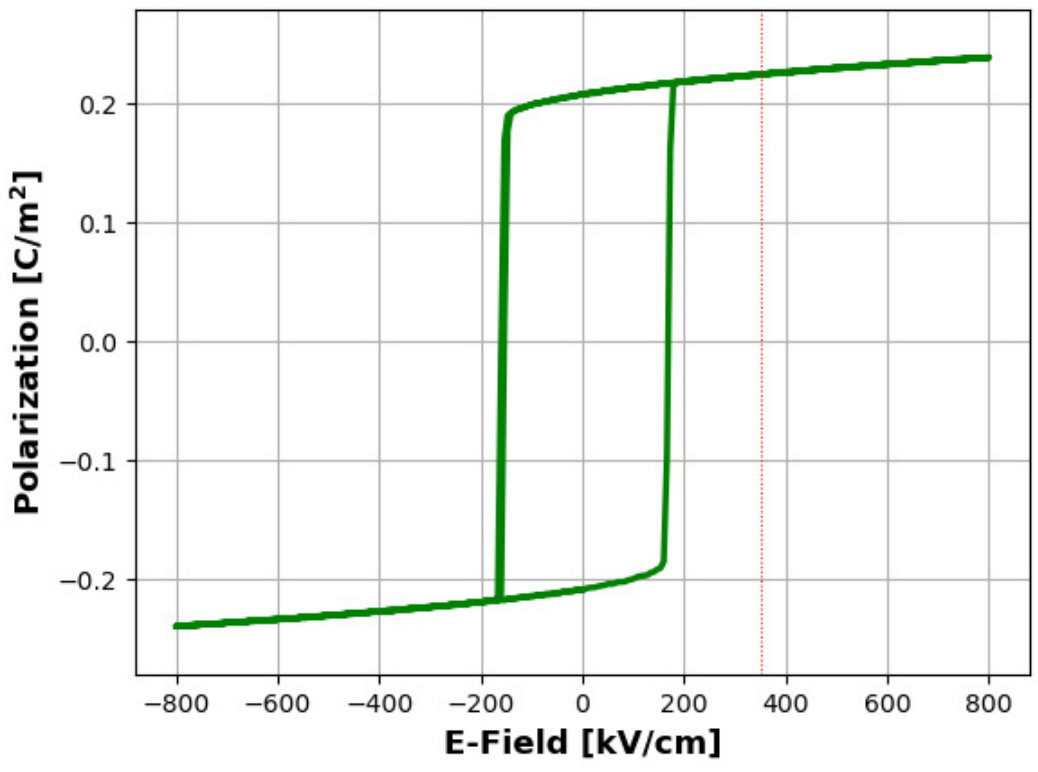

*Figure S6. Polarization–electric-field (P–E) hysteresis loop of the ordered 12.5% BZT superlattice computed from effective-Hamiltonian MD at T=293 K with the electric field applied along [111]. The polarization P is the supercell-averaged polarization projected onto the field direction (i.e., the [111] component). The vertical red line marks E=350 kV/cm, the local write field used in Section III.E to induce and stabilize the cylindrical nanodomains.*